\documentclass[aps,prl,twocolumn,superscriptaddress,amsfont,graphicx,nofootinbib,preprintnumbers]{revtex4-1}%
\UseRawInputEncoding
\usepackage{color,graphicx,epsfig}
\usepackage{ifpdf}
\usepackage{amsmath}
\usepackage{bm}
\usepackage[english]{babel}
\usepackage{amssymb}
\usepackage{braket}
\usepackage{hyperref}
\usepackage{setspace}

\usepackage{slashed}

\usepackage{changes}

\begin{document}

\title{New Strong Bounds on sub-GeV Dark Matter from Boosted and Migdal Effects}

\author{Victor V. Flambaum}
\email{v.flambaum@unsw.edu.au}
\affiliation{School of Physics, University of New South Wales, Sydney 2052, Australia}

\author{Liangliang Su}
\email{liangliangsu@njnu.edu.cn}
\affiliation{Department of Physics and Institute of Theoretical Physics, Nanjing Normal University, Nanjing, 210023, China}


\author{Lei Wu}
\email{leiwu@njnu.edu.cn}
\affiliation{Department of Physics and Institute of Theoretical Physics, Nanjing Normal University, Nanjing, 210023, China}


\author{Bin Zhu}
\email{zhubin@mail.nankai.edu.cn}
\affiliation{School of Physics, Yantai University, Yantai 264005, China}

\date{\today}

\begin{abstract}
Due to the low nuclear recoils, sub-GeV dark matter (DM) is usually beyond the sensitivity of the conventional DM direct detection experiments. The boosted and Migdal scattering mechanisms have been proposed as two new complementary avenues to search for light DM. In this work, we consider the momentum-transfer effect in the DM-nucleus scattering to derive the new bounds on sub-GeV DM for these two scenarios. We show that such an effect is sizable so that the existing bounds on the DM-nucleus scattering cross section can be improved significantly.
\end{abstract}
\pacs{Valid PACS appear here}
\maketitle


\section{Introduction}
The existence of dark matter has been supported by various cosmological and astrophysical observations. However, its properties, including the mass and interactions, are still elusive. As an attractive dark matter candidate, the Weakly Interacting Massive Particles (WIMPs) \cite{Lee:1977ua,Jungman:1995df} have been extensively studied in direct detection, indirect detection, and collider experiments~\cite{Roszkowski:2017nbc}. With unprecedented sensitivity, the current direct detections that measure the nuclear recoils have produced stringent limits on the models of the thermal relic DM in the sub-GeV mass range
generically requiring the presence of light mediators below the weak
scale. This motivates a broad investigation of sub-GeV dark matter particles~(see e.g.~\cite{Essig:2011nj,Hochberg:2015pha,Essig:2015cda,Essig:2017kqs,Knapen:2017xzo,Bertone:2018krk}).

As known, the sub-GeV DM from the standard halo usually produces too small energy deposited in the targets, $E_R \sim m^2_\chi v^2_\chi/ m_N \sim 0.1$ eV, to reach the threshold of the traditional noble liquid detectors. Direct detection of sub-GeV DM thus requires new approaches, both theoretically and experimentally. For example, the interactions between the high-energy primary cosmic rays (CRs) and the light DM may impart the energy to the DM, and thus change the speed distribution of the DM, which can produce a significant relativistic component. Then, these boosted light DM with sufficient kinetic energy can reach the sensitivity of conventional detectors~\cite{Bringmann:2018cvk,Cappiello:2018hsu,Ema:2018bih,Cappiello:2019qsw,Dent:2019krz, Bondarenko:2019vrb,Wang:2019jtk,Smirnov:2020zwf,Guo:2020drq, Ge:2020yuf,Cao:2020bwd,Guo:2020oum,Xia:2020wcp}. In addition, the boosted light DM can come from light meson decays produced in the inelastic collision of CRs with the atmosphere as well~\cite{Alvey:2019zaa,Su:2020zny}. By inheriting the energy of their parents, the light DM can move fast and generate a detectable scintillation or ionization signal in direct detection.

Besides resorting to astrophysical sources, the atomic ionization effect of materials can also be used to extend the sensitivity of the current direct detection. In the collision of the DM with the atom, due to the non-instantaneous movement of the electron cloud, the recoil atom is ionized, which deposits the energy inside the detector. This effect is the so-called Migdal scattering~\cite{Migdal:1939}. It can take place in a very low energy nuclear recoil, and extend the coverage of the low mass DM in the noble liquid targets and semiconductors~\cite{Vergados:2005dpd,Moustakidis:2005gx,Bernabei:2007jz,Ibe:2017yqa,Dolan:2017xbu,Bell:2019egg,Baxter:2019pnz,Essig:2019xkx,Liang:2019nnx,Nakamura:2020kex,GrillidiCortona:2020owp,Dey:2020sai,Liu:2020pat,Knapen:2020aky,Liang:2020ryg,He:2020sat}.

{In this work, we will study the momentum-transfer effects in the DM-nucleus scattering for the above two mechanisms of probing the light dark matter. Since models of the thermal relic DM in the sub-GeV mass range generically require the presence of light mediators below the weak scale, the new coupling to the Standard Model facilitates annihilation in the early universe and allows for the correct thermal relic abundance~(see, e.g.~\cite{Pospelov:2007mp,Arkani-Hamed:2008hhe,Feng:2008mu,Boehm:2003hm}). Besides, in models with the mediators, the momentum-transfer can significantly change the DM fluxes, ionization factor, and thus the expected recoil spectra. For example, when traveling through the earth, the DM particles will scatter with the nuclei of light elements, and be attenuated before reaching the detector. While if the DM can produce high momentum-transfer (about $q \gtrsim \mathcal{O}(10)$ MeV), such an attenuation effect can be mitigated because the DM-nucleus scattering cross section is highly suppressed by the nuclear form factor as $q \gtrsim \mathrm{O}(100)$ MeV. Thus, the DM fluxes in the high kinetic energy region around the detector will be affected. On the other hand, the ionization factor is the overlap integral between the wavefunctions of the particles involved in the collision, and thus should obey the orthogonality condition, i.e., the wave function of a bound electron has to be orthogonal to the wave function of the free electron because they are in different states. However, the orthogonality is not respected in the commonly used Hartree-Fock method. This will cause a numerical inaccuracy of the ionization factor{~\cite{An:2017ojc, Tan:2021nif}}.
Particularly in the low momentum transfer cases (about $q \lesssim \mathcal{O}(1)$ MeV), such as the Migdal scattering, this problem leads to incorrect results in the previous calculations. The solution to such a problem by subtracting contribution, which must vanish due to the orthogonality of the bound and continuum electron wave functions, significantly changes the exclusion limits compared to the previous study.}

\section{Model and Calculation}
The momentum-transfer effects can be important in dark matter models with light mediators. In our study, we consider a hadrophilic DM model~\cite{Batell:2018fqo}, in which a flavor-specific singlet scalar mediator $S$ couples to the up-type quarks $u$ and the Dirac dark matter $\chi$. The relevant effective interactions are given by,
\begin{eqnarray}
\mathcal{L}=i \bar{\chi}\left(\slashed D-m_{\chi}\right) \chi+\frac{1}{2} \partial_{\mu} S \partial^{\mu} S-\frac{1}{2} m_{S}^{2} S^{2} \nonumber \\ -\left(g_{\chi} S \bar{\chi}_{L} \chi_{R}+g_{u} S \bar{u}_{L} u_{R}+\mathrm{h.c.}\right)
\label{lag}
\end{eqnarray}
where $m_S$ and $m_\chi$ are the masses of mediator and DM, respectively. $g_\chi$ and $g_u$ are the couplings of mediator $S$ with the DM and up-quarks, respectively. The phenomenology of such a hadrophilic DM has been studied in Ref.~\cite{Batell:2018fqo,Su:2020zny}. With Eq.~\ref{lag}, we can obtain the DM-nucleon scattering cross section,
\begin{eqnarray}
\sigma_{\mathbf n}=\bar{\sigma}_{\mathbf n}\left|F_{\mathrm{DM}}(q)\right|^{2},    
\label{cx}
\end{eqnarray}
where the momentum-independent cross section $\bar{\sigma}_{\mathbf n}$ is defined by
\begin{eqnarray}
\bar{\sigma}_{\mathbf n} \equiv \mu_{\mathbf n}^{2} \frac{\overline{\left|\mathcal{M}_{\mathbf n}\left(q=q_{0}\right)\right|^{2}}}{\left(16 \pi m_{\chi}^{2} m_{\mathbf n}^{2}\right)}=\frac{g_{\mathbf n}^2 g_{\chi }^2 \mu_{\mathbf n}^2}{\pi \left( q^2_0+m_S^2\right)^2},
\label{sigmaN}    
\end{eqnarray}
and the DM form factor $F_{\mathrm{DM}}(q)$ is defined by,
\begin{eqnarray}
\left|F_{\mathrm{DM}}(q)\right|^{2} &\equiv& \frac{\overline{\left|\mathcal{M}_{\mathbf n}(q)\right|^{2}}}{\left|\mathcal{M}_{\mathbf n}\left(q=q_{0}\right)\right|^{2}} \nonumber \\
&=& \frac{\left(4m_{\mathbf n}^2+q^2\right) \left(q_0^2+m_S^2\right)^2 \left(4 m_{\chi}^2+q^2\right)}{ 16 m_{\mathbf n}^2 m_{\chi }^2 \left(m_S^2+q^2\right)^2}.
\label{formDM}
\end{eqnarray}
Here $\mu_{\mathbf n} = m_\chi m_{\mathbf n}/(m_\chi + m_{\mathbf n})$ is the DM and nucleon reduced mass and $g_\mathbf{n}=0.014 g_u (m_p/m_u)$ is the effective coupling between the scalar mediator and a nucleon. $q=p^\prime_{\chi} - p_{\chi}$ is the momentum-transfer and $q_0=\alpha m_e$ is the reference momentum-transfer. In general, the DM form factor is a function of the momentum-transfer and depends on the specific model. Note that the light DM in the halo usually lacks enough energy to produce the nuclear recoil signals in the conventional direct detections. In order to solve this, one can use the sophisticated atomic effect, such as Migdal scattering, which can generate the ionized electrons via the low momentum-transfer from the DM to the nucleus. On the other hand, the light DM can obtain the extra kinetic energy through the up-scattering with the CRs or from the decay of the heavy particles, and thus produce the nucleus recoil by the high momentum-transfer. In the following, we will demonstrate that such high or low momentum-transfer effects in the DM-nucleus scattering process including the attenuation and ionization can significantly change the existing exclusion limits.

\subsection{High momentum-transfer via boosted effect}

As an example of the high momentum-transfer{(about $q \gtrsim \mathcal{O}(10)$ MeV)} light DM, we study the atmospheric dark matter (ADM), which arises from the decays of the mesons produced in the inelastic collision of high-energy CRs with the atmosphere on Earth~\cite{Alvey:2019zaa, Su:2020zny}. In our work, we focus on the following production process of the ADM,
\begin{eqnarray}
p+N \to \eta \to \pi S(\to \chi\bar{\chi}),   
\end{eqnarray}
where $p$ is the proton in the CRs and $N$ is the nitrogen in the atmosphere. $\eta$ and $\pi$ are the SM light mesons. In our calculation, we only include the contribution of the proton in the flux of CRs from the local interstellar (LIS). We follow Ref.~\cite{Boschini:2017fxq} and take the same parameters to calculate the differential flux of the proton in the CRs as $\mathrm{d} \phi_p/\mathrm{d} T_p = 4 \pi (\mathrm{d} R/\mathrm{d} T_p) (\mathrm{d} I/\mathrm{d} R)$, where the $\mathrm{d} I/\mathrm{d} R$ is the differential density and $R$ is the particle's rigidity.

The differential recoil rate of the DM-nucleus scattering per unit target mass per unit time per keV can be calculated by,
\begin{eqnarray}
    \frac{\mathrm{d} R}{\mathrm{d} E_R} &= N_T\int \frac{\mathrm{d} \sigma}{\mathrm{d} E_R}\frac{\mathrm{d} \Phi_\chi}{\mathrm{d} E_{\chi}} \mathrm{d} E_{\chi} \nonumber \\
    &=N_T n_{\text {halo }} \dfrac{{\rm d} \langle \sigma v \rangle}{{\rm d} E_{R}},
\label{drate}
\end{eqnarray}
with
\begin{eqnarray}
    \frac{{\rm d} \langle \sigma v \rangle}{{\rm d} E_{R}} &= \dfrac{\sigma_{\mathbf n} m_N}{2\mu^2_{\mathbf n}} \left[ f_p Z + f_n (A-Z)\right]^2 \nonumber \\
    & \times |F_N(q)|^2 \eta(E^{min}_\chi),
\label{sigmav} 
\end{eqnarray}
where the DM-nucleon scattering cross section $\sigma_{\mathbf n}$ is given in Eq.~\ref{cx}. $N_T=4.2\times 10^{27}$ is number density of Xenon per tonne. $n_{\text {halo }}=\rho_\chi^{\text{local}}/m_\chi$ is the halo DM number density. The recoil energy $E_R=q^2/2m_N$ is the function of the momentum transfer $q$ and the nucleus mass of the target $m_N$. $Z$ is the atomic number and $A$ is the mass number of atoms. $f_{p,n}$ are dimensionless couplings between DM and proton/neutron. We assume the iso-spin conservation and take $f_p = f_n$ in our calculations. Besides, we use the Helm nucleus form factor $F_N(q)$ in Eq.~\ref{sigmav}, which is given by~\cite{Duda:2006uk},
\begin{equation}
    \left|F_{N}\right|^2 = \left(\frac{3 j_{1}(q R_1)}{q R_1}\right)^2 e^{- q^2 s^2}
\label{nuclearform}
\end{equation}
with
\begin{equation}
    j_1(x) = \frac{\sin x}{x^2}-\frac{\cos x}{x}
\end{equation}
being the spherical Bessel function of the first kind. $R_1$ and $s$ are the effective nuclear radius and skin thickness, respectively. Their values for the Helm form factor that can reproduce the numerical fourier transform of a Two-Parameter Fermi distribution are taken as
\begin{eqnarray}
R_1(x) = \sqrt{c^2+\frac{7}{3}\pi^2 a^2 -5s^2},\;\; s = 0.9 \; \rm{fm}    
\end{eqnarray}
where $a=0.52 \;\rm{fm}$ and $c=(1.23 A^{1/3} -0.60)\;\rm{fm}$.

The inverse speed function $\eta(E^{min}_\chi)$ in Eq.~\ref{sigmav} is defined as
\begin{eqnarray}
\eta\left(E^{min}_{\chi}\right)=\int_{E^{min}_{\chi}} \mathrm{d} E_{\chi} n_{\text {halo }}^{-1} \frac{m_{\chi}^{2}}{p_\chi^2} \frac{\mathrm{d} \Phi_{\chi}}{\mathrm{d} E_{\chi}},
\label{velocity}
\end{eqnarray}
where $p_{\chi}$ is the DM momentum. The minimum DM energy $E_{\chi}^{min}$ is given by
\begin{equation}
\begin{aligned}
    E_{\chi}^{min} &=m_\chi+ \left(\frac{E_R}{2}-m_{\chi}\right)\\
    &\times \left(1\pm \sqrt{1+\frac{2 E_R}{m_N}\frac{(m_\chi+m_N)^2}{(2m_{\chi}-E_R)^2}}\right),
\end{aligned}
\end{equation}
where the plus and minus sign apply for $E_R>2 m_{\chi}$ and $E_R<2 m_{\chi}$,respectively. The differential flux of the ADM $\mathrm{d}\Phi_\chi/\mathrm{d}E_{\chi}$ in Eq.~\ref{velocity} at the detector can be calculated by,
\begin{eqnarray}
    \frac{\mathrm{d} \Phi_{\chi}}{\mathrm{d} E_{\chi}} = \frac{\mathrm{d} \Phi_{\chi}}{\mathrm{d} T_{\chi}} =\frac{\mathrm{d} T_{\chi}^0}{\mathrm{d} T_{\chi}} \frac{\mathrm{d} \Phi_\chi}{ \mathrm{d} T_{\chi}^0},
\end{eqnarray}
where $\mathrm{d} \Phi_\chi/\mathrm{d} T_{\chi}^{0}$ and $T_{\chi}^0$ are the differential flux of the ADM and the kinetic energy of the ADM at the Earth's surface, respectively. Before reaching the detector, the ADM will lose energy and thus be attenuated when it travels through the Earth. The dependence of the kinetic energy $T_{\chi}$ on the distance $l$ between the DM production point and the detector can be evaluated by, 
\begin{eqnarray}
&&\frac{\mathrm{d} T_{\mathrm{\chi}}}{\mathrm{d} l}=-\sum_{i} n_{i}(\vec{r}) \frac{\bar{\sigma}_{\mathbf n} m_{i}}{2 \mu_{\mathbf n}^{2}}\frac{m_{\chi}^2}{T_{\chi}^2+2 m_{\chi} T_{\chi}}A_i^{2} \nonumber \\
&& \times \int_{0}^{E_{R}^{\max }}   \left|F_{\mathrm{DM}}(E_R)\right|^{2}\left|F_{N}(E_R)\right|^{2} E_{R} \mathrm{d} E_{R}
\label{eq:dTchidl}
\end{eqnarray}
where $n_i(\vec{r})$ is the number density of the different medium particles $i$ of Earth at position $\vec{r}$ and $E_{R}$ is the DM recoil energy, respectively. The detailed discussions of the attenuation and form factors are given in the supplemental material. Considering the existing bounds from the meson decays, the Big Bang nucleosynthesis and direct detections~\cite{Adler:2002hy,Adler:2004hp,Adler:2008zza,Artamonov:2009sz,Aguilar-Arevalo:2018wea}, we take $g_u=10^{-5}$ and $m_S=300$ MeV as our benchmark point, which corresponds to $\mathrm{BR}(\eta \to \pi \chi \bar{\chi}) \simeq 1\times10^{-5}$ in our numerical calculations.

\begin{figure}[ht]
  \centering
  \includegraphics[width=7cm,height=6cm]{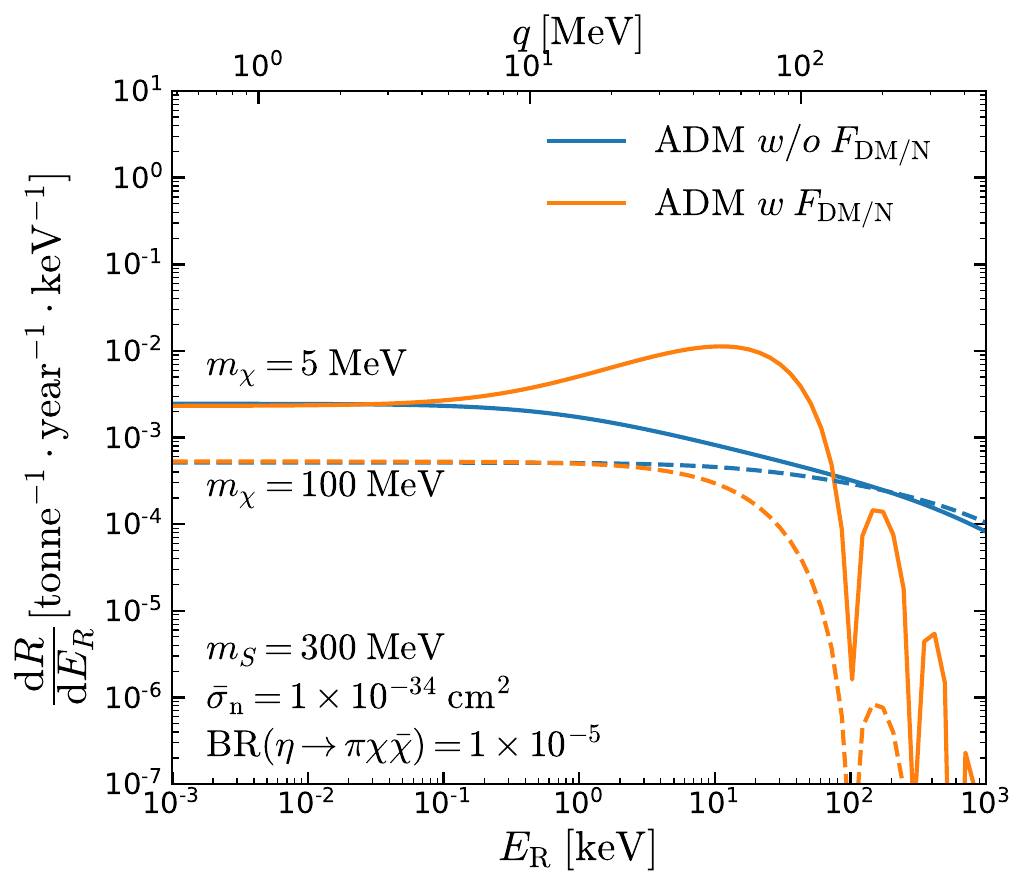}
  \caption{The differential recoil rate $\mathrm{d}R/\mathrm{d}E_R$ of the ADM-nucleus scattering process, where $E_R$ and $q$ stands for the nucleus recoil and the DM momentum transfer. We compare the results with and without $F_{\mathrm{DM}}$ and $F_N$. The solid and dashed lines correspond to two benchmark DM masses $m_\chi=$ 5 MeV and 100 MeV, respectively.}
  \label{fig:rate1}
\end{figure}
In Fig.~\ref{fig:rate1}, we present the impact of the momentum transfer on the differential recoil rates of the ADM-nucleus scattering in Eq.~\ref{drate}. We find that the nucleus recoil $E_R$ produced by the ADM can reach the threshold of the nucleus recoil signal ($E_R \sim 1$ keV) in Xenon experiments. The lighter DM has a larger differential recoil rate because it has a larger kinematically allowed region. On the other hand, it should be noted that the momentum-transfer effect in the DM form factor $F_{\mathrm{DM}}$ for lighter DM, such as $m_\chi=5$ MeV, leads to a sizable enhancement of the differential recoil rate in the range of $0.1 \lesssim E_R \lesssim 100$ keV. For the higher $E_R$, the differential rate will be highly suppressed by the nucleus form factor $F_N$.

\begin{figure}[ht!]
\centering
\includegraphics[width=7cm,height=6cm]{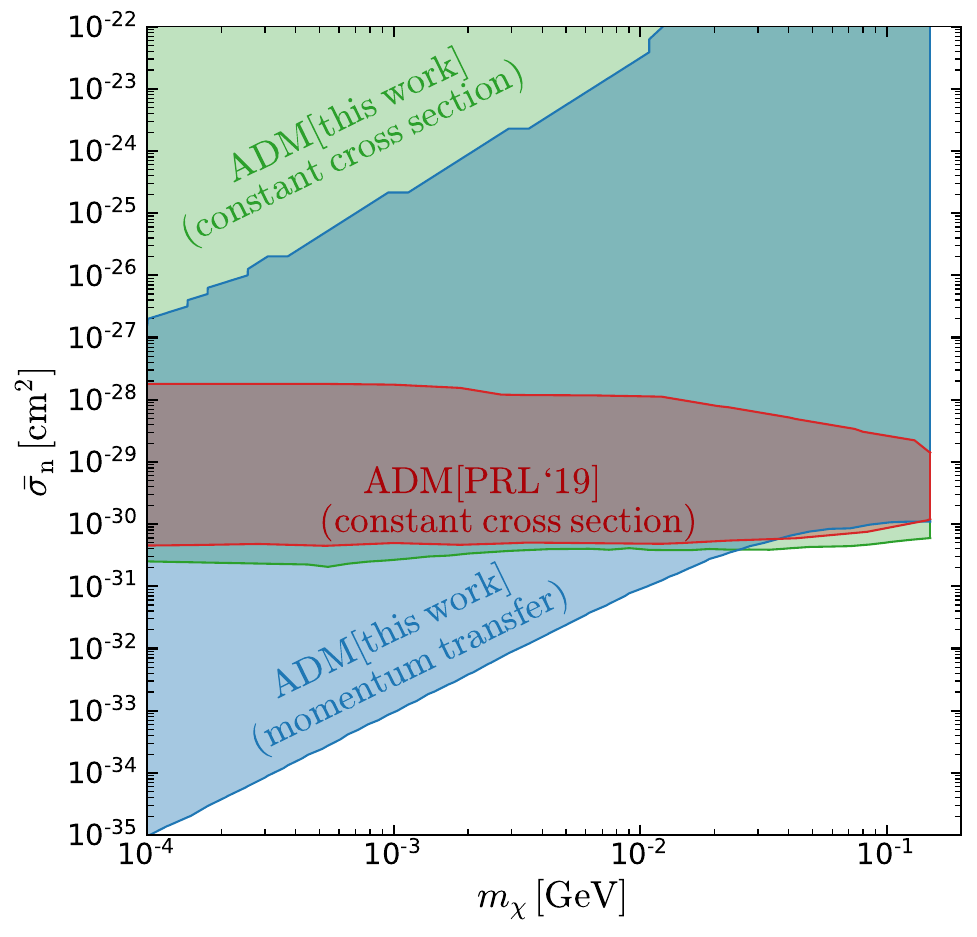}
\caption{The Xenon1T exclusion limits on the momentum-independent ADM-nucleon scattering cross section $\bar{\sigma}_{\mathbf n}$ in Eq.~\ref{sigmaN} with or without the momentum-transfer corrections. As a comparison, the previous result in Ref.~\cite{Alvey:2019zaa} that is rescaled by the branching ration $Br(\eta \to \pi \bar{\chi}\chi)=10^{-5}$ is also shown.}
\label{fig:ADM_limit}
\end{figure}
In Fig.~\ref{fig:ADM_limit}, we show the exclusion limits on the momentum-independent ADM-nucleon scattering cross section $\bar{\sigma}_{\mathbf n}$ by using the Xenon1T data~\cite{XENON:2018voc,Bringmann:2018cvk} (see supplemental material). In Ref.~\cite{Alvey:2019zaa}, the upper limit $\bar{\sigma}_n \sim 10^{-28}$ $\mathrm{cm}^{2}$ was obtained for a constant cross section case with the DM form factor $F_{\mathrm{DM}}=1$ in Eq.~\ref{cx}. The nucleus form factor $F_N$ in Eq.~\ref{nuclearform} was only considered in the ADM-nucleus scattering process inside the detector. However, we find that such a bound can be greatly improved by including the nucleus form factor $F_{N}$ in the attenuation because the ADM-nucleus scattering cross section is highly suppressed by $F_N$ in the large $q$ region. This suppression effect will decrease the energy loss of the ADM during its travels through the Earth so that the ADM with the large scattering cross section can still reach the detector and thus be excluded. {And the green upper limit can reach about ${\cal O}(10^{-19})$ cm$^2$. Note that such large scattering cross sections are also limited by the cosmological and astrophysical observations, such as BBN and CMB\cite{Gluscevic:2017ywp,Batell:2018fqo,Xu:2018efh,Ooba:2019erm}.} Besides, if further including the momentum-dependent DM form factor $F_{\mathrm{DM}}$ in Eq.~\ref{formDM}, we can obtain a much stronger limit on the ADM-nucleus scattering cross section $\bar{\sigma}_{\mathbf n}$ in our mass range due to $1/m_{\chi}^2$ enhancement of $F_{\mathrm{DM}}$ in $q \gtrsim m_{\chi}$ range, as shown in Fig.~\ref{fig:formfactor} of the supplemental material.

\subsection{Low momentum-transfer via atomic effect}

When the momentum transfer in the DM-nucleus scattering is small{(about $q \lesssim \mathcal{O}(1)$ MeV)}, such as the sub-GeV DM in the halo, the Migdal mechanism provides a way to detect light DM via the electron ionization. {ADM is not being considered in this section and halo DM is being considered.} The differential event rate of the ionized electron can be written as,
\begin{eqnarray}
      \frac{\mathrm{d} R_{i o n}}{\mathrm{d} E_{e}}&=&N_T\int \frac{\mathrm{d}^2 \sigma }{\mathrm{d} E_R \mathrm{d} E_e}\frac{\mathrm{d} \Phi_\chi}{\mathrm{d} E_{\chi}} \mathrm{d} E_{\chi}\mathrm{d} E_R \nonumber \\
&=&N_{T} n_{\rm halo} \sum_{n l} \frac{\mathrm{d} \left\langle\sigma_{ion}^{n l} v\right\rangle}{\mathrm{d} E_{e}},
\label{mrate}
\end{eqnarray}
with
\begin{eqnarray}
&&\frac{\mathrm{d}\left\langle\sigma^{nl}_{ion} v\right\rangle}{\mathrm{d}  E_{e}}
=\frac{[f_p Z+f_n (A-Z)]^2}{8 \mu_{\mathbf n}^{2} E_e} \int_{q_{\min}} \mathrm{d} q q \nonumber \\
&&\times \sigma_{\mathbf n} \left|F_{N}(q)\right|^{2} \left|f^{n l}_{ion}\left(p_{e}, q_{e}\right)\right|^{2} \eta\left(E^{min}_{\chi}\right),
\label{ionizationcx}
\end{eqnarray}
where $n$ and $l$ stand for the principal quantum number and the orbital angular momentum quantum number of the electron in the Xenon atom, respectively. $E_e$ and $p_e$ are the energy and momentum of the ionized electron, respectively. $q$ and $q_e$ are the momentum transfers of the nucleus and the electron, respectively, which are related by $q_e \simeq (m_e/m_N) q \sim 10^{-6} (m_\chi v)$ in the Migdal scattering. The minimal value of $q$ should be able to ionize the electrons in the first shell. With the energy conservation, we can have the minimal value of the DM energy $E^{min}_{\chi} \simeq (q+|E_{nl}|+ E_e)/2$, where $E_{nl}$ is the binding energy. Note that the Eq.~\ref{velocity} is a generalized will give the usual form, $\eta=\int_{v_{\chi}^{min}} f(\mathbf v)/v \mathrm{d}^3 \mathbf v$ for the halo DM, which is discussed in the supplemental material.

The ionization factor $\left|f^{n l}_{ion}\left(p_{e}, q_{e}\right)\right|^2$ in Eq.~\ref{ionizationcx} is proportional to the possibility that electrons in an atom are ionized during a nuclear recoil in the inelastic collisions between the dark matter and the atom~\cite{Essig:2019xkx}. For simplicity, we will not include the secondary contributions from the de-excitation electron and the $X$-rays in this work. In our study, we compute the ionization factor by numerically solving the Schrodinger equation with a certain potential. We adopt the Roothaan-Hartree-Fock (RHF) ground state wave functions for the initial electron state wave function. Thus, the ionization form factor can be written as~\cite{Essig:2012yx}
\begin{eqnarray}
\left|f^{nl}_{ion}(p_e,q_e)\right|^2 &=  \frac{4 p_e^3}{(2 \pi)^3} \sum_{\ell^{\prime}=0}^{\infty} \sum_{L =|\ell-\ell^{\prime}|}^{\ell+\ell} (2\ell+1) \nonumber \\
    & \times (2\ell^{\prime}+1) (2L+1) \left(\begin{array}{ccc}
\ell & \ell^{\prime} & L \nonumber \\
0 & 0 & 0
\end{array}\right)^{2} \\
& \times \int_{0}^{\infty} \mathrm{d} r r^{2} j_{L}(q_e r) R_{p_e \ell^{\prime}}^{*}(r) R_{n \ell}(r), 
\end{eqnarray}   
where $R_{n\ell, p_e \ell^{\prime}}$ are the radial part of initial and final electron state wave function. $j_L$ is the spherical Bessel function of the first kind.

It should be noted that the orthogonality condition of those wave functions in the low momentum transfer $q$ region will be violated because of the nonzero results of radial integral of the $L=0$ term $\int \mathrm{d} r r^{2} j_{0}(q_e r) R_{p_e \ell^{\prime}}^{*}(r) R_{n \ell}(r)$ in the numerical calculations. In Ref.~\cite{Essig:2019xkx}, they deal with this problem technically by neglecting the $L=0$ term and using dipole approximation at the low $q_e$ region, {which we will refer to as ``without Subtract" scheme in this work.} However, this method will lose the contribution of $L=0$ term at a moderate value of $q_e$. To solve such a problem, we modify this radial integral as $\int \mathrm{d} r r^{2} (j_{0}(q_e r)-1) R_{p_e \ell^{\prime}}^{*}(r) R_{n \ell}(r)$ to guarantee the orthogonality when $q_e r \ll 1$, {i.e., the ``Subtract" scheme.} We include the contributions from the electron configurations of $4s$, $4p$, $4d$, $5s$ and $5p$. Since the momentum transfer is small, we will not include the relativistic correction to ionization factor~\cite{Roberts:2016xfw}. The numerical calculation is performed within the framework of \textsf{DarkARC}~\cite{Catena:2019gfa}. 

\begin{figure}[ht]
  \centering
  \includegraphics[width=7cm,height=6cm]{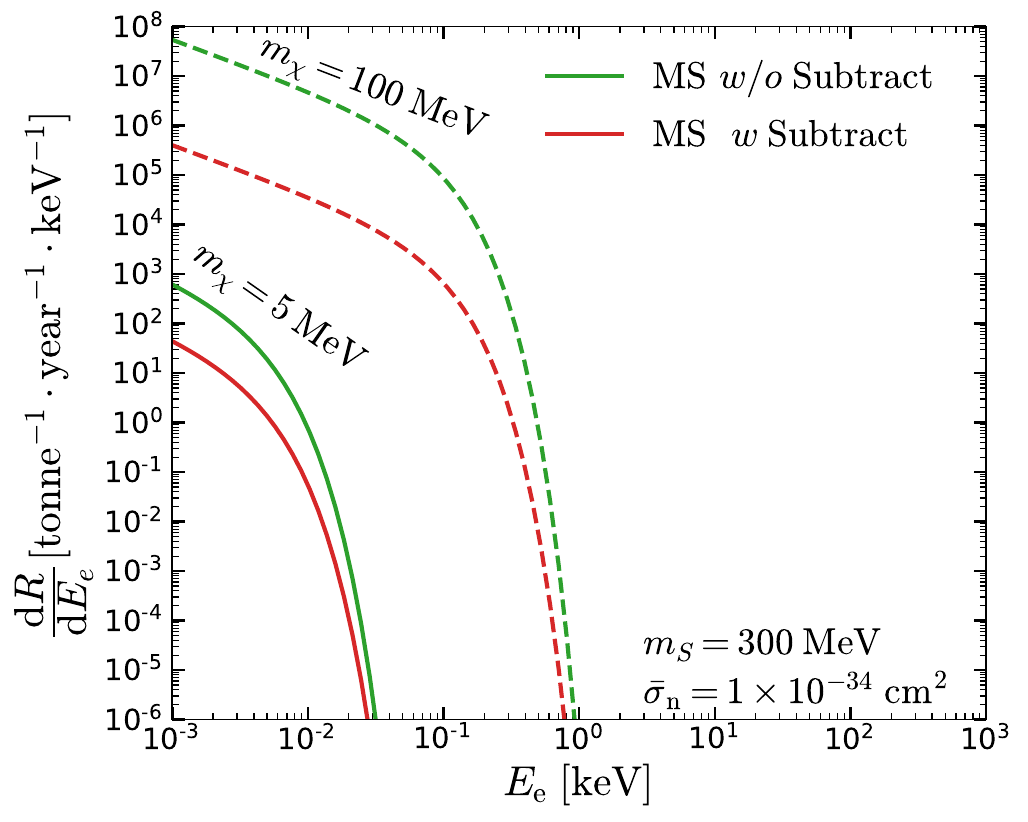}
  \caption{The differential rates of the DM-nucleus Migdal scattering process, where $E_e$ stands for the electron recoil. The contributions of the momentum dependent $F_{\mathrm{DM}}$ in Eq.~\ref{formDM} and $F_{\mathrm{DM}}$~in Eq.~\ref{nuclearform} are included. We compare the results with or without the orthogonality subtraction. The solid and dashed lines correspond to DM mass $m_\chi=$ 5 MeV and 100 MeV, respectively.}\label{fig:rate2}
\end{figure}
In Fig.~\ref{fig:rate2}, we show the differential rates $\mathrm{d}R/\mathrm{d}E_e$ of the DM-nucleus Migdal scattering process in Eq.~\ref{mrate}. We find that the Migdal scattering can produce sizable events of the ionized electron in the low energy $E_e$. With the increase of the DM mass, the differential rate becomes large because of the enhancement of the ionization factor $f_{ion}$ in the high $q_e$ range (see supplemental material). We also note that the results with the orthogonality subtraction in the low $E_e$ region greatly differ from that without the orthogonality subtraction, which demonstrates the contribution of the $L=0$ term to the Migdal scattering cannot be neglected.

\begin{figure}[ht!]
\centering
\includegraphics[width=7cm,height=6cm]{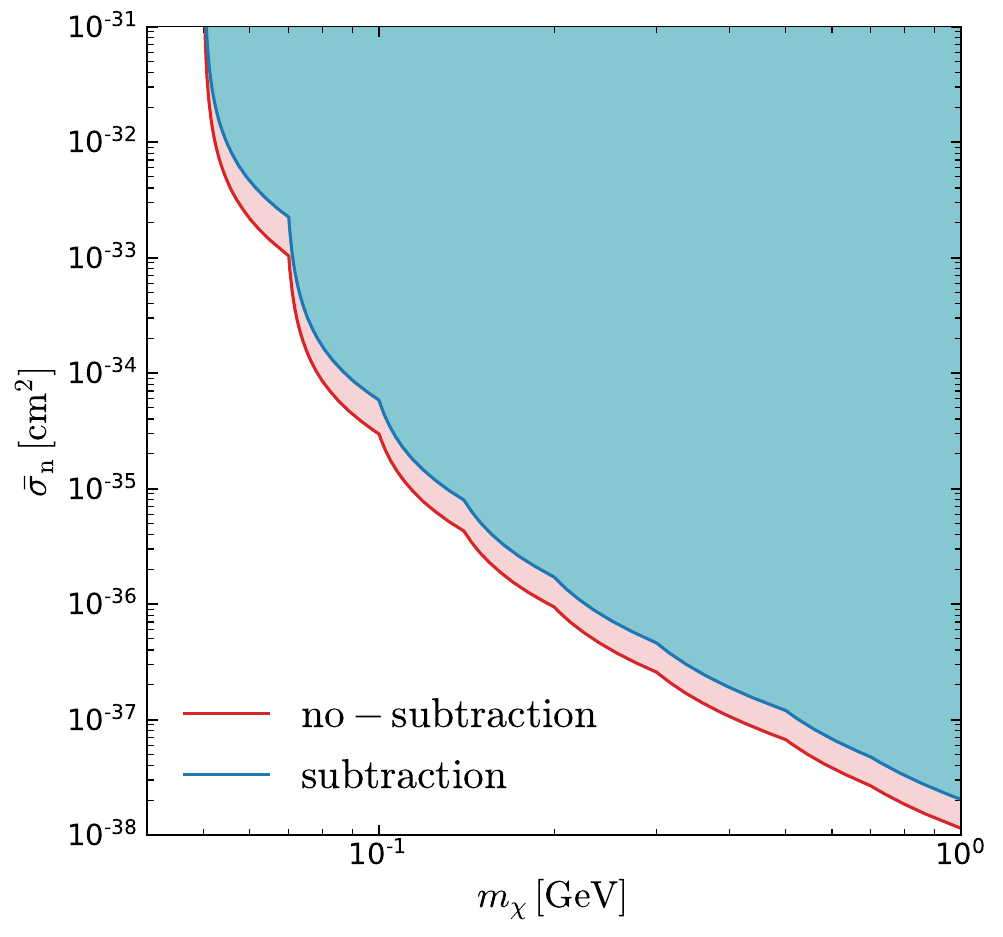}
\caption{The exclusion limits on the momentum-independent DM-nucleon scattering cross section $\bar{\sigma}_{\mathbf n}$ with the orthogonality subtraction through the Migdal scattering. As a comparison, we also show the results without subtraction.}
\label{fig:Migdal_limit}
\end{figure}
In Fig.~\ref{fig:Migdal_limit}, we employ the Xenon1T ionization data~\cite{Aprile:2019xxb,xenon:collaboration} to derive the exclusion limits on the momentum-independent DM-nucleon scattering cross section $\bar{\sigma}_{\mathbf n}$ through the Migdal scattering. As a comparison, we show the results with or without the orthogonality subtraction. {It can be seen that the exclusion limit without the subtraction is about two times larger than that with the subtraction. }This is because the ionization factor becomes smaller after the subtraction. On the other hand, we also checked and find that the bound of the ADM with the Migdal effect is weaker than that of the ADM because of the suppression of the ionization factor. In other words, once the light DM can obtain enough kinetic energy, the current nucleus recoil data can provide a stronger limit than the ionization data.


Finally, we discuss the relic density of ADM in our interest of mass range, i.e. $m_{\chi}<m_{S}$, in which $\chi$ will annihilate directly to the SM quarks via an s-channel mediator. The cross section of annihilation is given by 
\begin{equation}
\sigma v_{\mathrm{rel}}({\chi \bar{\chi} \rightarrow q\bar{q}})=\frac{\left.g_{\chi}^{2} m_{\chi} v_{\mathrm{rel}}^{2} \Gamma_{S}\right|_{m_{S}=2 m_{\chi}}}{2\left(\left(m_{S}^{2}-4 m_{\chi}^{2}\right)^{2}+m_{S}^{2} \Gamma_{S}^{2}\right)},
\end{equation}
which is suppressed by velocity. $\left.\Gamma_{S}\right|_{m_{S}=2 m_{\chi}}$ is the decay width of the scalar at $m_{S}=2 m_{\chi}$. The freeze-out is guaranteed that $\Gamma_S>H[m_{S}]$, which in turn favors relatively larger coupling $g_u$. Thus DM is in chemical equilibrium with the early universe bath as long as it contains quarks. After the QCD phase transition takes place, DM pairs could still annihilate into two pions. Therefore achieving the correct relic density via thermal freeze-out would require $m_{\chi}>m_{\pi}$. For lighter DM, we can allow for a small coupling of the mediator to neutrinos to achieve the correct relic abundance via thermal freeze-out~\cite{Berlin:2017ftj}.

\section{acknowledgments}

LW is supported by the National Natural Science Foundation of China (NNSFC) under grant No. 12275134. BZ is supported by the National Natural Science Foundation of China (NNSFC) under grant No. 12275232, and by the Natural Science Foundation of Shandong Province under grant ZR2018QA007. VF is supported by the Australian Research Council Grants No. DP190100974 and DP200100150.  We are grateful to Ben Roberts for a useful discussion.

\bibliography{refs}

\begin{thebibliography}{65}
\expandafter\ifx\csname natexlab\endcsname\relax\def\natexlab#1{#1}\fi
\expandafter\ifx\csname bibnamefont\endcsname\relax
  \def\bibnamefont#1{#1}\fi
\expandafter\ifx\csname bibfnamefont\endcsname\relax
  \def\bibfnamefont#1{#1}\fi
\expandafter\ifx\csname citenamefont\endcsname\relax
  \def\citenamefont#1{#1}\fi
\expandafter\ifx\csname url\endcsname\relax
  \def\url#1{\texttt{#1}}\fi
\expandafter\ifx\csname urlprefix\endcsname\relax\def\urlprefix{URL }\fi
\providecommand{\bibinfo}[2]{#2}
\providecommand{\eprint}[2][]{\url{#2}}

\bibitem[{\citenamefont{Lee and Weinberg}(1977)}]{Lee:1977ua}
\bibinfo{author}{\bibfnamefont{B.~W.} \bibnamefont{Lee}} \bibnamefont{and}
  \bibinfo{author}{\bibfnamefont{S.}~\bibnamefont{Weinberg}},
  \bibinfo{journal}{Phys. Rev. Lett.} \textbf{\bibinfo{volume}{39}},
  \bibinfo{pages}{165} (\bibinfo{year}{1977}).

\bibitem[{\citenamefont{Jungman et~al.}(1996)\citenamefont{Jungman,
  Kamionkowski, and Griest}}]{Jungman:1995df}
\bibinfo{author}{\bibfnamefont{G.}~\bibnamefont{Jungman}},
  \bibinfo{author}{\bibfnamefont{M.}~\bibnamefont{Kamionkowski}},
  \bibnamefont{and} \bibinfo{author}{\bibfnamefont{K.}~\bibnamefont{Griest}},
  \bibinfo{journal}{Phys. Rept.} \textbf{\bibinfo{volume}{267}},
  \bibinfo{pages}{195} (\bibinfo{year}{1996}), \eprint{hep-ph/9506380}.

\bibitem[{\citenamefont{Roszkowski et~al.}(2018)\citenamefont{Roszkowski,
  Sessolo, and Trojanowski}}]{Roszkowski:2017nbc}
\bibinfo{author}{\bibfnamefont{L.}~\bibnamefont{Roszkowski}},
  \bibinfo{author}{\bibfnamefont{E.~M.} \bibnamefont{Sessolo}},
  \bibnamefont{and}
  \bibinfo{author}{\bibfnamefont{S.}~\bibnamefont{Trojanowski}},
  \bibinfo{journal}{Rept. Prog. Phys.} \textbf{\bibinfo{volume}{81}},
  \bibinfo{pages}{066201} (\bibinfo{year}{2018}), \eprint{1707.06277}.

\bibitem[{\citenamefont{Essig et~al.}(2012{\natexlab{a}})\citenamefont{Essig,
  Mardon, and Volansky}}]{Essig:2011nj}
\bibinfo{author}{\bibfnamefont{R.}~\bibnamefont{Essig}},
  \bibinfo{author}{\bibfnamefont{J.}~\bibnamefont{Mardon}}, \bibnamefont{and}
  \bibinfo{author}{\bibfnamefont{T.}~\bibnamefont{Volansky}},
  \bibinfo{journal}{Phys. Rev. D} \textbf{\bibinfo{volume}{85}},
  \bibinfo{pages}{076007} (\bibinfo{year}{2012}{\natexlab{a}}),
  \eprint{1108.5383}.

\bibitem[{\citenamefont{Hochberg et~al.}(2016)\citenamefont{Hochberg, Zhao, and
  Zurek}}]{Hochberg:2015pha}
\bibinfo{author}{\bibfnamefont{Y.}~\bibnamefont{Hochberg}},
  \bibinfo{author}{\bibfnamefont{Y.}~\bibnamefont{Zhao}}, \bibnamefont{and}
  \bibinfo{author}{\bibfnamefont{K.~M.} \bibnamefont{Zurek}},
  \bibinfo{journal}{Phys. Rev. Lett.} \textbf{\bibinfo{volume}{116}},
  \bibinfo{pages}{011301} (\bibinfo{year}{2016}), \eprint{1504.07237}.

\bibitem[{\citenamefont{Essig et~al.}(2016)\citenamefont{Essig,
  Fernandez-Serra, Mardon, Soto, Volansky, and Yu}}]{Essig:2015cda}
\bibinfo{author}{\bibfnamefont{R.}~\bibnamefont{Essig}},
  \bibinfo{author}{\bibfnamefont{M.}~\bibnamefont{Fernandez-Serra}},
  \bibinfo{author}{\bibfnamefont{J.}~\bibnamefont{Mardon}},
  \bibinfo{author}{\bibfnamefont{A.}~\bibnamefont{Soto}},
  \bibinfo{author}{\bibfnamefont{T.}~\bibnamefont{Volansky}}, \bibnamefont{and}
  \bibinfo{author}{\bibfnamefont{T.-T.} \bibnamefont{Yu}},
  \bibinfo{journal}{JHEP} \textbf{\bibinfo{volume}{05}}, \bibinfo{pages}{046}
  (\bibinfo{year}{2016}), \eprint{1509.01598}.

\bibitem[{\citenamefont{Essig et~al.}(2017)\citenamefont{Essig, Volansky, and
  Yu}}]{Essig:2017kqs}
\bibinfo{author}{\bibfnamefont{R.}~\bibnamefont{Essig}},
  \bibinfo{author}{\bibfnamefont{T.}~\bibnamefont{Volansky}}, \bibnamefont{and}
  \bibinfo{author}{\bibfnamefont{T.-T.} \bibnamefont{Yu}},
  \bibinfo{journal}{Phys. Rev. D} \textbf{\bibinfo{volume}{96}},
  \bibinfo{pages}{043017} (\bibinfo{year}{2017}), \eprint{1703.00910}.

\bibitem[{\citenamefont{Knapen et~al.}(2017)\citenamefont{Knapen, Lin, and
  Zurek}}]{Knapen:2017xzo}
\bibinfo{author}{\bibfnamefont{S.}~\bibnamefont{Knapen}},
  \bibinfo{author}{\bibfnamefont{T.}~\bibnamefont{Lin}}, \bibnamefont{and}
  \bibinfo{author}{\bibfnamefont{K.~M.} \bibnamefont{Zurek}},
  \bibinfo{journal}{Phys. Rev. D} \textbf{\bibinfo{volume}{96}},
  \bibinfo{pages}{115021} (\bibinfo{year}{2017}), \eprint{1709.07882}.

\bibitem[{\citenamefont{Bertone and Tait}(2018)}]{Bertone:2018krk}
\bibinfo{author}{\bibfnamefont{G.}~\bibnamefont{Bertone}} \bibnamefont{and}
  \bibinfo{author}{\bibfnamefont{M.}~\bibnamefont{Tait}, \bibfnamefont{Tim}},
  \bibinfo{journal}{Nature} \textbf{\bibinfo{volume}{562}}, \bibinfo{pages}{51}
  (\bibinfo{year}{2018}), \eprint{1810.01668}.

\bibitem[{\citenamefont{Bringmann and Pospelov}(2019)}]{Bringmann:2018cvk}
\bibinfo{author}{\bibfnamefont{T.}~\bibnamefont{Bringmann}} \bibnamefont{and}
  \bibinfo{author}{\bibfnamefont{M.}~\bibnamefont{Pospelov}},
  \bibinfo{journal}{Phys. Rev. Lett.} \textbf{\bibinfo{volume}{122}},
  \bibinfo{pages}{171801} (\bibinfo{year}{2019}), \eprint{1810.10543}.

\bibitem[{\citenamefont{Cappiello et~al.}(2019)\citenamefont{Cappiello, Ng, and
  Beacom}}]{Cappiello:2018hsu}
\bibinfo{author}{\bibfnamefont{C.~V.} \bibnamefont{Cappiello}},
  \bibinfo{author}{\bibfnamefont{K.~C.} \bibnamefont{Ng}}, \bibnamefont{and}
  \bibinfo{author}{\bibfnamefont{J.~F.} \bibnamefont{Beacom}},
  \bibinfo{journal}{Phys. Rev. D} \textbf{\bibinfo{volume}{99}},
  \bibinfo{pages}{063004} (\bibinfo{year}{2019}), \eprint{1810.07705}.

\bibitem[{\citenamefont{Ema et~al.}(2019)\citenamefont{Ema, Sala, and
  Sato}}]{Ema:2018bih}
\bibinfo{author}{\bibfnamefont{Y.}~\bibnamefont{Ema}},
  \bibinfo{author}{\bibfnamefont{F.}~\bibnamefont{Sala}}, \bibnamefont{and}
  \bibinfo{author}{\bibfnamefont{R.}~\bibnamefont{Sato}},
  \bibinfo{journal}{Phys. Rev. Lett.} \textbf{\bibinfo{volume}{122}},
  \bibinfo{pages}{181802} (\bibinfo{year}{2019}), \eprint{1811.00520}.

\bibitem[{\citenamefont{Cappiello and Beacom}(2019)}]{Cappiello:2019qsw}
\bibinfo{author}{\bibfnamefont{C.~V.} \bibnamefont{Cappiello}}
  \bibnamefont{and} \bibinfo{author}{\bibfnamefont{J.~F.}
  \bibnamefont{Beacom}}, \bibinfo{journal}{Phys. Rev. D}
  \textbf{\bibinfo{volume}{100}}, \bibinfo{pages}{103011}
  (\bibinfo{year}{2019}), \bibinfo{note}{[Erratum: Phys.Rev.D 104, 069901
  (2021)]}, \eprint{1906.11283}.

\bibitem[{\citenamefont{Dent et~al.}(2020)\citenamefont{Dent, Dutta, Newstead,
  and Shoemaker}}]{Dent:2019krz}
\bibinfo{author}{\bibfnamefont{J.~B.} \bibnamefont{Dent}},
  \bibinfo{author}{\bibfnamefont{B.}~\bibnamefont{Dutta}},
  \bibinfo{author}{\bibfnamefont{J.~L.} \bibnamefont{Newstead}},
  \bibnamefont{and} \bibinfo{author}{\bibfnamefont{I.~M.}
  \bibnamefont{Shoemaker}}, \bibinfo{journal}{Phys. Rev. D}
  \textbf{\bibinfo{volume}{101}}, \bibinfo{pages}{116007}
  (\bibinfo{year}{2020}), \eprint{1907.03782}.

\bibitem[{\citenamefont{Bondarenko et~al.}(2020)\citenamefont{Bondarenko,
  Boyarsky, Bringmann, Hufnagel, Schmidt-Hoberg, and
  Sokolenko}}]{Bondarenko:2019vrb}
\bibinfo{author}{\bibfnamefont{K.}~\bibnamefont{Bondarenko}},
  \bibinfo{author}{\bibfnamefont{A.}~\bibnamefont{Boyarsky}},
  \bibinfo{author}{\bibfnamefont{T.}~\bibnamefont{Bringmann}},
  \bibinfo{author}{\bibfnamefont{M.}~\bibnamefont{Hufnagel}},
  \bibinfo{author}{\bibfnamefont{K.}~\bibnamefont{Schmidt-Hoberg}},
  \bibnamefont{and}
  \bibinfo{author}{\bibfnamefont{A.}~\bibnamefont{Sokolenko}},
  \bibinfo{journal}{JHEP} \textbf{\bibinfo{volume}{03}}, \bibinfo{pages}{118}
  (\bibinfo{year}{2020}), \eprint{1909.08632}.

\bibitem[{\citenamefont{Wang et~al.}(2020)\citenamefont{Wang, Wu, Yang, Zhou,
  and Zhu}}]{Wang:2019jtk}
\bibinfo{author}{\bibfnamefont{W.}~\bibnamefont{Wang}},
  \bibinfo{author}{\bibfnamefont{L.}~\bibnamefont{Wu}},
  \bibinfo{author}{\bibfnamefont{J.~M.} \bibnamefont{Yang}},
  \bibinfo{author}{\bibfnamefont{H.}~\bibnamefont{Zhou}}, \bibnamefont{and}
  \bibinfo{author}{\bibfnamefont{B.}~\bibnamefont{Zhu}},
  \bibinfo{journal}{JHEP} \textbf{\bibinfo{volume}{12}}, \bibinfo{pages}{072}
  (\bibinfo{year}{2020}), \bibinfo{note}{[Erratum: JHEP 02, 052 (2021)]},
  \eprint{1912.09904}.

\bibitem[{\citenamefont{Smirnov and Beacom}(2020)}]{Smirnov:2020zwf}
\bibinfo{author}{\bibfnamefont{J.}~\bibnamefont{Smirnov}} \bibnamefont{and}
  \bibinfo{author}{\bibfnamefont{J.~F.} \bibnamefont{Beacom}}
  (\bibinfo{year}{2020}), \eprint{2002.04038}.

\bibitem[{\citenamefont{Guo et~al.}(2020{\natexlab{a}})\citenamefont{Guo, Tsai,
  and Wu}}]{Guo:2020drq}
\bibinfo{author}{\bibfnamefont{G.}~\bibnamefont{Guo}},
  \bibinfo{author}{\bibfnamefont{Y.-L.~S.} \bibnamefont{Tsai}},
  \bibnamefont{and} \bibinfo{author}{\bibfnamefont{M.-R.} \bibnamefont{Wu}},
  \bibinfo{journal}{JCAP} \textbf{\bibinfo{volume}{10}}, \bibinfo{pages}{049}
  (\bibinfo{year}{2020}{\natexlab{a}}), \eprint{2004.03161}.

\bibitem[{\citenamefont{Ge et~al.}(2020)\citenamefont{Ge, Liu, Yuan, and
  Zhou}}]{Ge:2020yuf}
\bibinfo{author}{\bibfnamefont{S.-F.} \bibnamefont{Ge}},
  \bibinfo{author}{\bibfnamefont{J.-L.} \bibnamefont{Liu}},
  \bibinfo{author}{\bibfnamefont{Q.}~\bibnamefont{Yuan}}, \bibnamefont{and}
  \bibinfo{author}{\bibfnamefont{N.}~\bibnamefont{Zhou}}
  (\bibinfo{year}{2020}), \eprint{2005.09480}.

\bibitem[{\citenamefont{Cao et~al.}(2020)\citenamefont{Cao, Ding, and
  Xiang}}]{Cao:2020bwd}
\bibinfo{author}{\bibfnamefont{Q.-H.} \bibnamefont{Cao}},
  \bibinfo{author}{\bibfnamefont{R.}~\bibnamefont{Ding}}, \bibnamefont{and}
  \bibinfo{author}{\bibfnamefont{Q.-F.} \bibnamefont{Xiang}}
  (\bibinfo{year}{2020}), \eprint{2006.12767}.

\bibitem[{\citenamefont{Guo et~al.}(2020{\natexlab{b}})\citenamefont{Guo, Tsai,
  Wu, and Yuan}}]{Guo:2020oum}
\bibinfo{author}{\bibfnamefont{G.}~\bibnamefont{Guo}},
  \bibinfo{author}{\bibfnamefont{Y.-L.~S.} \bibnamefont{Tsai}},
  \bibinfo{author}{\bibfnamefont{M.-R.} \bibnamefont{Wu}}, \bibnamefont{and}
  \bibinfo{author}{\bibfnamefont{Q.}~\bibnamefont{Yuan}},
  \bibinfo{journal}{Phys. Rev. D} \textbf{\bibinfo{volume}{102}},
  \bibinfo{pages}{103004} (\bibinfo{year}{2020}{\natexlab{b}}),
  \eprint{2008.12137}.

\bibitem[{\citenamefont{Xia et~al.}(2020)\citenamefont{Xia, Xu, and
  Zhou}}]{Xia:2020wcp}
\bibinfo{author}{\bibfnamefont{C.}~\bibnamefont{Xia}},
  \bibinfo{author}{\bibfnamefont{Y.-H.} \bibnamefont{Xu}}, \bibnamefont{and}
  \bibinfo{author}{\bibfnamefont{Y.-F.} \bibnamefont{Zhou}}
  (\bibinfo{year}{2020}), \eprint{2009.00353}.

\bibitem[{\citenamefont{Alvey et~al.}(2019)\citenamefont{Alvey, Campos,
  Fairbairn, and You}}]{Alvey:2019zaa}
\bibinfo{author}{\bibfnamefont{J.}~\bibnamefont{Alvey}},
  \bibinfo{author}{\bibfnamefont{M.}~\bibnamefont{Campos}},
  \bibinfo{author}{\bibfnamefont{M.}~\bibnamefont{Fairbairn}},
  \bibnamefont{and} \bibinfo{author}{\bibfnamefont{T.}~\bibnamefont{You}},
  \bibinfo{journal}{Phys. Rev. Lett.} \textbf{\bibinfo{volume}{123}},
  \bibinfo{pages}{261802} (\bibinfo{year}{2019}), \eprint{1905.05776}.

\bibitem[{\citenamefont{Su et~al.}(2020)\citenamefont{Su, Wang, Wu, Yang, and
  Zhu}}]{Su:2020zny}
\bibinfo{author}{\bibfnamefont{L.}~\bibnamefont{Su}},
  \bibinfo{author}{\bibfnamefont{W.}~\bibnamefont{Wang}},
  \bibinfo{author}{\bibfnamefont{L.}~\bibnamefont{Wu}},
  \bibinfo{author}{\bibfnamefont{J.~M.} \bibnamefont{Yang}}, \bibnamefont{and}
  \bibinfo{author}{\bibfnamefont{B.}~\bibnamefont{Zhu}} (\bibinfo{year}{2020}),
  \eprint{2006.11837}.

\bibitem[{\citenamefont{Migdal}(1939)}]{Migdal:1939}
\bibinfo{author}{\bibfnamefont{A.}~\bibnamefont{Migdal}},
  \bibinfo{journal}{Sov.Phys.JETP} \textbf{\bibinfo{volume}{9}},
  \bibinfo{pages}{1163} (\bibinfo{year}{1939}).

\bibitem[{\citenamefont{Vergados and Ejiri}(2005)}]{Vergados:2005dpd}
\bibinfo{author}{\bibfnamefont{J.~D.} \bibnamefont{Vergados}} \bibnamefont{and}
  \bibinfo{author}{\bibfnamefont{H.}~\bibnamefont{Ejiri}},
  \bibinfo{journal}{Phys. Lett. B} \textbf{\bibinfo{volume}{606}},
  \bibinfo{pages}{313} (\bibinfo{year}{2005}), \eprint{hep-ph/0401151}.

\bibitem[{\citenamefont{Moustakidis et~al.}(2005)\citenamefont{Moustakidis,
  Vergados, and Ejiri}}]{Moustakidis:2005gx}
\bibinfo{author}{\bibfnamefont{C.~C.} \bibnamefont{Moustakidis}},
  \bibinfo{author}{\bibfnamefont{J.~D.} \bibnamefont{Vergados}},
  \bibnamefont{and} \bibinfo{author}{\bibfnamefont{H.}~\bibnamefont{Ejiri}},
  \bibinfo{journal}{Nucl. Phys. B} \textbf{\bibinfo{volume}{727}},
  \bibinfo{pages}{406} (\bibinfo{year}{2005}), \eprint{hep-ph/0507123}.

\bibitem[{\citenamefont{Bernabei et~al.}(2007)}]{Bernabei:2007jz}
\bibinfo{author}{\bibfnamefont{R.}~\bibnamefont{Bernabei}}
  \bibnamefont{et~al.}, \bibinfo{journal}{Int. J. Mod. Phys. A}
  \textbf{\bibinfo{volume}{22}}, \bibinfo{pages}{3155} (\bibinfo{year}{2007}),
  \eprint{0706.1421}.

\bibitem[{\citenamefont{Ibe et~al.}(2018)\citenamefont{Ibe, Nakano, Shoji, and
  Suzuki}}]{Ibe:2017yqa}
\bibinfo{author}{\bibfnamefont{M.}~\bibnamefont{Ibe}},
  \bibinfo{author}{\bibfnamefont{W.}~\bibnamefont{Nakano}},
  \bibinfo{author}{\bibfnamefont{Y.}~\bibnamefont{Shoji}}, \bibnamefont{and}
  \bibinfo{author}{\bibfnamefont{K.}~\bibnamefont{Suzuki}},
  \bibinfo{journal}{JHEP} \textbf{\bibinfo{volume}{03}}, \bibinfo{pages}{194}
  (\bibinfo{year}{2018}), \eprint{1707.07258}.

\bibitem[{\citenamefont{Dolan et~al.}(2018)\citenamefont{Dolan, Kahlhoefer, and
  McCabe}}]{Dolan:2017xbu}
\bibinfo{author}{\bibfnamefont{M.~J.} \bibnamefont{Dolan}},
  \bibinfo{author}{\bibfnamefont{F.}~\bibnamefont{Kahlhoefer}},
  \bibnamefont{and} \bibinfo{author}{\bibfnamefont{C.}~\bibnamefont{McCabe}},
  \bibinfo{journal}{Phys. Rev. Lett.} \textbf{\bibinfo{volume}{121}},
  \bibinfo{pages}{101801} (\bibinfo{year}{2018}), \eprint{1711.09906}.

\bibitem[{\citenamefont{Bell et~al.}(2020)\citenamefont{Bell, Dent, Newstead,
  Sabharwal, and Weiler}}]{Bell:2019egg}
\bibinfo{author}{\bibfnamefont{N.~F.} \bibnamefont{Bell}},
  \bibinfo{author}{\bibfnamefont{J.~B.} \bibnamefont{Dent}},
  \bibinfo{author}{\bibfnamefont{J.~L.} \bibnamefont{Newstead}},
  \bibinfo{author}{\bibfnamefont{S.}~\bibnamefont{Sabharwal}},
  \bibnamefont{and} \bibinfo{author}{\bibfnamefont{T.~J.}
  \bibnamefont{Weiler}}, \bibinfo{journal}{Phys. Rev. D}
  \textbf{\bibinfo{volume}{101}}, \bibinfo{pages}{015012}
  (\bibinfo{year}{2020}), \eprint{1905.00046}.

\bibitem[{\citenamefont{Baxter et~al.}(2020)\citenamefont{Baxter, Kahn, and
  Krnjaic}}]{Baxter:2019pnz}
\bibinfo{author}{\bibfnamefont{D.}~\bibnamefont{Baxter}},
  \bibinfo{author}{\bibfnamefont{Y.}~\bibnamefont{Kahn}}, \bibnamefont{and}
  \bibinfo{author}{\bibfnamefont{G.}~\bibnamefont{Krnjaic}},
  \bibinfo{journal}{Phys. Rev. D} \textbf{\bibinfo{volume}{101}},
  \bibinfo{pages}{076014} (\bibinfo{year}{2020}), \eprint{1908.00012}.

\bibitem[{\citenamefont{Essig et~al.}(2020)\citenamefont{Essig, Pradler,
  Sholapurkar, and Yu}}]{Essig:2019xkx}
\bibinfo{author}{\bibfnamefont{R.}~\bibnamefont{Essig}},
  \bibinfo{author}{\bibfnamefont{J.}~\bibnamefont{Pradler}},
  \bibinfo{author}{\bibfnamefont{M.}~\bibnamefont{Sholapurkar}},
  \bibnamefont{and} \bibinfo{author}{\bibfnamefont{T.-T.} \bibnamefont{Yu}},
  \bibinfo{journal}{Phys. Rev. Lett.} \textbf{\bibinfo{volume}{124}},
  \bibinfo{pages}{021801} (\bibinfo{year}{2020}), \eprint{1908.10881}.

\bibitem[{\citenamefont{Liang et~al.}(2020{\natexlab{a}})\citenamefont{Liang,
  Zhang, Zheng, and Zhang}}]{Liang:2019nnx}
\bibinfo{author}{\bibfnamefont{Z.-L.} \bibnamefont{Liang}},
  \bibinfo{author}{\bibfnamefont{L.}~\bibnamefont{Zhang}},
  \bibinfo{author}{\bibfnamefont{F.}~\bibnamefont{Zheng}}, \bibnamefont{and}
  \bibinfo{author}{\bibfnamefont{P.}~\bibnamefont{Zhang}},
  \bibinfo{journal}{Phys. Rev. D} \textbf{\bibinfo{volume}{102}},
  \bibinfo{pages}{043007} (\bibinfo{year}{2020}{\natexlab{a}}),
  \eprint{1912.13484}.

\bibitem[{\citenamefont{Nakamura et~al.}(2020)\citenamefont{Nakamura, Miuchi,
  Kazama, Shoji, Ibe, and Nakano}}]{Nakamura:2020kex}
\bibinfo{author}{\bibfnamefont{K.~D.} \bibnamefont{Nakamura}},
  \bibinfo{author}{\bibfnamefont{K.}~\bibnamefont{Miuchi}},
  \bibinfo{author}{\bibfnamefont{S.}~\bibnamefont{Kazama}},
  \bibinfo{author}{\bibfnamefont{Y.}~\bibnamefont{Shoji}},
  \bibinfo{author}{\bibfnamefont{M.}~\bibnamefont{Ibe}}, \bibnamefont{and}
  \bibinfo{author}{\bibfnamefont{W.}~\bibnamefont{Nakano}}
  (\bibinfo{year}{2020}), \eprint{2009.05939}.

\bibitem[{\citenamefont{Grilli~di Cortona et~al.}(2020)\citenamefont{Grilli~di
  Cortona, Messina, and Piacentini}}]{GrillidiCortona:2020owp}
\bibinfo{author}{\bibfnamefont{G.}~\bibnamefont{Grilli~di Cortona}},
  \bibinfo{author}{\bibfnamefont{A.}~\bibnamefont{Messina}}, \bibnamefont{and}
  \bibinfo{author}{\bibfnamefont{S.}~\bibnamefont{Piacentini}},
  \bibinfo{journal}{JHEP} \textbf{\bibinfo{volume}{11}}, \bibinfo{pages}{034}
  (\bibinfo{year}{2020}), \eprint{2006.02453}.

\bibitem[{\citenamefont{Dey et~al.}(2020)\citenamefont{Dey, Maity, and
  Ray}}]{Dey:2020sai}
\bibinfo{author}{\bibfnamefont{U.~K.} \bibnamefont{Dey}},
  \bibinfo{author}{\bibfnamefont{T.~N.} \bibnamefont{Maity}}, \bibnamefont{and}
  \bibinfo{author}{\bibfnamefont{T.~S.} \bibnamefont{Ray}},
  \bibinfo{journal}{Phys. Lett. B} \textbf{\bibinfo{volume}{811}},
  \bibinfo{pages}{135900} (\bibinfo{year}{2020}), \eprint{2006.12529}.

\bibitem[{\citenamefont{Liu et~al.}(2020)\citenamefont{Liu, Wu, Chi, and
  Chen}}]{Liu:2020pat}
\bibinfo{author}{\bibfnamefont{C.-P.} \bibnamefont{Liu}},
  \bibinfo{author}{\bibfnamefont{C.-P.} \bibnamefont{Wu}},
  \bibinfo{author}{\bibfnamefont{H.-C.} \bibnamefont{Chi}}, \bibnamefont{and}
  \bibinfo{author}{\bibfnamefont{J.-W.} \bibnamefont{Chen}}
  (\bibinfo{year}{2020}), \eprint{2007.10965}.

\bibitem[{\citenamefont{Knapen et~al.}(2020)\citenamefont{Knapen, Kozaczuk, and
  Lin}}]{Knapen:2020aky}
\bibinfo{author}{\bibfnamefont{S.}~\bibnamefont{Knapen}},
  \bibinfo{author}{\bibfnamefont{J.}~\bibnamefont{Kozaczuk}}, \bibnamefont{and}
  \bibinfo{author}{\bibfnamefont{T.}~\bibnamefont{Lin}} (\bibinfo{year}{2020}),
  \eprint{2011.09496}.

\bibitem[{\citenamefont{Liang et~al.}(2020{\natexlab{b}})\citenamefont{Liang,
  Mo, Zheng, and Zhang}}]{Liang:2020ryg}
\bibinfo{author}{\bibfnamefont{Z.-L.} \bibnamefont{Liang}},
  \bibinfo{author}{\bibfnamefont{C.}~\bibnamefont{Mo}},
  \bibinfo{author}{\bibfnamefont{F.}~\bibnamefont{Zheng}}, \bibnamefont{and}
  \bibinfo{author}{\bibfnamefont{P.}~\bibnamefont{Zhang}}
  (\bibinfo{year}{2020}{\natexlab{b}}), \eprint{2011.13352}.

\bibitem[{\citenamefont{He et~al.}(2021)\citenamefont{He, Wang, and
  Zheng}}]{He:2020sat}
\bibinfo{author}{\bibfnamefont{H.-J.} \bibnamefont{He}},
  \bibinfo{author}{\bibfnamefont{Y.-C.} \bibnamefont{Wang}}, \bibnamefont{and}
  \bibinfo{author}{\bibfnamefont{J.}~\bibnamefont{Zheng}},
  \bibinfo{journal}{Phys. Rev. D} \textbf{\bibinfo{volume}{104}},
  \bibinfo{pages}{115033} (\bibinfo{year}{2021}), \eprint{2012.05891}.

\bibitem[{\citenamefont{Pospelov et~al.}(2008)\citenamefont{Pospelov, Ritz, and
  Voloshin}}]{Pospelov:2007mp}
\bibinfo{author}{\bibfnamefont{M.}~\bibnamefont{Pospelov}},
  \bibinfo{author}{\bibfnamefont{A.}~\bibnamefont{Ritz}}, \bibnamefont{and}
  \bibinfo{author}{\bibfnamefont{M.~B.} \bibnamefont{Voloshin}},
  \bibinfo{journal}{Phys. Lett. B} \textbf{\bibinfo{volume}{662}},
  \bibinfo{pages}{53} (\bibinfo{year}{2008}), \eprint{0711.4866}.

\bibitem[{\citenamefont{Arkani-Hamed et~al.}(2009)\citenamefont{Arkani-Hamed,
  Finkbeiner, Slatyer, and Weiner}}]{Arkani-Hamed:2008hhe}
\bibinfo{author}{\bibfnamefont{N.}~\bibnamefont{Arkani-Hamed}},
  \bibinfo{author}{\bibfnamefont{D.~P.} \bibnamefont{Finkbeiner}},
  \bibinfo{author}{\bibfnamefont{T.~R.} \bibnamefont{Slatyer}},
  \bibnamefont{and} \bibinfo{author}{\bibfnamefont{N.}~\bibnamefont{Weiner}},
  \bibinfo{journal}{Phys. Rev. D} \textbf{\bibinfo{volume}{79}},
  \bibinfo{pages}{015014} (\bibinfo{year}{2009}), \eprint{0810.0713}.

\bibitem[{\citenamefont{Feng et~al.}(2008)\citenamefont{Feng, Tu, and
  Yu}}]{Feng:2008mu}
\bibinfo{author}{\bibfnamefont{J.~L.} \bibnamefont{Feng}},
  \bibinfo{author}{\bibfnamefont{H.}~\bibnamefont{Tu}}, \bibnamefont{and}
  \bibinfo{author}{\bibfnamefont{H.-B.} \bibnamefont{Yu}},
  \bibinfo{journal}{JCAP} \textbf{\bibinfo{volume}{10}}, \bibinfo{pages}{043}
  (\bibinfo{year}{2008}), \eprint{0808.2318}.

\bibitem[{\citenamefont{Boehm and Fayet}(2004)}]{Boehm:2003hm}
\bibinfo{author}{\bibfnamefont{C.}~\bibnamefont{Boehm}} \bibnamefont{and}
  \bibinfo{author}{\bibfnamefont{P.}~\bibnamefont{Fayet}},
  \bibinfo{journal}{Nucl. Phys. B} \textbf{\bibinfo{volume}{683}},
  \bibinfo{pages}{219} (\bibinfo{year}{2004}), \eprint{hep-ph/0305261}.

\bibitem[{\citenamefont{An et~al.}(2018)\citenamefont{An, Pospelov, Pradler,
  and Ritz}}]{An:2017ojc}
\bibinfo{author}{\bibfnamefont{H.}~\bibnamefont{An}},
  \bibinfo{author}{\bibfnamefont{M.}~\bibnamefont{Pospelov}},
  \bibinfo{author}{\bibfnamefont{J.}~\bibnamefont{Pradler}}, \bibnamefont{and}
  \bibinfo{author}{\bibfnamefont{A.}~\bibnamefont{Ritz}},
  \bibinfo{journal}{Phys. Rev. Lett.} \textbf{\bibinfo{volume}{120}},
  \bibinfo{pages}{141801} (\bibinfo{year}{2018}), \bibinfo{note}{[Erratum:
  Phys.Rev.Lett. 121, 259903 (2018)]}, \eprint{1708.03642}.

\bibitem[{\citenamefont{Tan et~al.}(2021)\citenamefont{Tan, Derevianko, Dzuba,
  and Flambaum}}]{Tan:2021nif}
\bibinfo{author}{\bibfnamefont{H.~B.~T.} \bibnamefont{Tan}},
  \bibinfo{author}{\bibfnamefont{A.}~\bibnamefont{Derevianko}},
  \bibinfo{author}{\bibfnamefont{V.~A.} \bibnamefont{Dzuba}}, \bibnamefont{and}
  \bibinfo{author}{\bibfnamefont{V.~V.} \bibnamefont{Flambaum}},
  \bibinfo{journal}{Phys. Rev. Lett.} \textbf{\bibinfo{volume}{127}},
  \bibinfo{pages}{081301} (\bibinfo{year}{2021}), \eprint{2105.08296}.

\bibitem[{\citenamefont{Batell et~al.}(2019)\citenamefont{Batell, Freitas,
  Ismail, and Mckeen}}]{Batell:2018fqo}
\bibinfo{author}{\bibfnamefont{B.}~\bibnamefont{Batell}},
  \bibinfo{author}{\bibfnamefont{A.}~\bibnamefont{Freitas}},
  \bibinfo{author}{\bibfnamefont{A.}~\bibnamefont{Ismail}}, \bibnamefont{and}
  \bibinfo{author}{\bibfnamefont{D.}~\bibnamefont{Mckeen}},
  \bibinfo{journal}{Phys. Rev. D} \textbf{\bibinfo{volume}{100}},
  \bibinfo{pages}{095020} (\bibinfo{year}{2019}), \eprint{1812.05103}.

\bibitem[{\citenamefont{Boschini et~al.}(2017)}]{Boschini:2017fxq}
\bibinfo{author}{\bibfnamefont{M.}~\bibnamefont{Boschini}}
  \bibnamefont{et~al.}, \bibinfo{journal}{Astrophys. J.}
  \textbf{\bibinfo{volume}{840}}, \bibinfo{pages}{115} (\bibinfo{year}{2017}),
  \eprint{1704.06337}.

\bibitem[{\citenamefont{Duda et~al.}(2007)\citenamefont{Duda, Kemper, and
  Gondolo}}]{Duda:2006uk}
\bibinfo{author}{\bibfnamefont{G.}~\bibnamefont{Duda}},
  \bibinfo{author}{\bibfnamefont{A.}~\bibnamefont{Kemper}}, \bibnamefont{and}
  \bibinfo{author}{\bibfnamefont{P.}~\bibnamefont{Gondolo}},
  \bibinfo{journal}{JCAP} \textbf{\bibinfo{volume}{04}}, \bibinfo{pages}{012}
  (\bibinfo{year}{2007}), \eprint{hep-ph/0608035}.

\bibitem[{\citenamefont{Adler et~al.}(2002)}]{Adler:2002hy}
\bibinfo{author}{\bibfnamefont{S.}~\bibnamefont{Adler}} \bibnamefont{et~al.}
  (\bibinfo{collaboration}{E787}), \bibinfo{journal}{Phys. Lett. B}
  \textbf{\bibinfo{volume}{537}}, \bibinfo{pages}{211} (\bibinfo{year}{2002}),
  \eprint{hep-ex/0201037}.

\bibitem[{\citenamefont{Adler et~al.}(2004)}]{Adler:2004hp}
\bibinfo{author}{\bibfnamefont{S.}~\bibnamefont{Adler}} \bibnamefont{et~al.}
  (\bibinfo{collaboration}{E787}), \bibinfo{journal}{Phys. Rev. D}
  \textbf{\bibinfo{volume}{70}}, \bibinfo{pages}{037102}
  (\bibinfo{year}{2004}), \eprint{hep-ex/0403034}.

\bibitem[{\citenamefont{Adler et~al.}(2008)}]{Adler:2008zza}
\bibinfo{author}{\bibfnamefont{S.}~\bibnamefont{Adler}} \bibnamefont{et~al.}
  (\bibinfo{collaboration}{E949, E787}), \bibinfo{journal}{Phys. Rev. D}
  \textbf{\bibinfo{volume}{77}}, \bibinfo{pages}{052003}
  (\bibinfo{year}{2008}), \eprint{0709.1000}.

\bibitem[{\citenamefont{Artamonov et~al.}(2009)}]{Artamonov:2009sz}
\bibinfo{author}{\bibfnamefont{A.}~\bibnamefont{Artamonov}}
  \bibnamefont{et~al.} (\bibinfo{collaboration}{BNL-E949}),
  \bibinfo{journal}{Phys. Rev. D} \textbf{\bibinfo{volume}{79}},
  \bibinfo{pages}{092004} (\bibinfo{year}{2009}), \eprint{0903.0030}.

\bibitem[{\citenamefont{Aguilar-Arevalo
  et~al.}(2018)}]{Aguilar-Arevalo:2018wea}
\bibinfo{author}{\bibfnamefont{A.}~\bibnamefont{Aguilar-Arevalo}}
  \bibnamefont{et~al.} (\bibinfo{collaboration}{MiniBooNE DM}),
  \bibinfo{journal}{Phys. Rev. D} \textbf{\bibinfo{volume}{98}},
  \bibinfo{pages}{112004} (\bibinfo{year}{2018}), \eprint{1807.06137}.

\bibitem[{\citenamefont{Aprile et~al.}(2018)}]{XENON:2018voc}
\bibinfo{author}{\bibfnamefont{E.}~\bibnamefont{Aprile}} \bibnamefont{et~al.}
  (\bibinfo{collaboration}{XENON}), \bibinfo{journal}{Phys. Rev. Lett.}
  \textbf{\bibinfo{volume}{121}}, \bibinfo{pages}{111302}
  (\bibinfo{year}{2018}), \eprint{1805.12562}.

\bibitem[{\citenamefont{Gluscevic and Boddy}(2018)}]{Gluscevic:2017ywp}
\bibinfo{author}{\bibfnamefont{V.}~\bibnamefont{Gluscevic}} \bibnamefont{and}
  \bibinfo{author}{\bibfnamefont{K.~K.} \bibnamefont{Boddy}},
  \bibinfo{journal}{Phys. Rev. Lett.} \textbf{\bibinfo{volume}{121}},
  \bibinfo{pages}{081301} (\bibinfo{year}{2018}), \eprint{1712.07133}.

\bibitem[{\citenamefont{Xu et~al.}(2018)\citenamefont{Xu, Dvorkin, and
  Chael}}]{Xu:2018efh}
\bibinfo{author}{\bibfnamefont{W.~L.} \bibnamefont{Xu}},
  \bibinfo{author}{\bibfnamefont{C.}~\bibnamefont{Dvorkin}}, \bibnamefont{and}
  \bibinfo{author}{\bibfnamefont{A.}~\bibnamefont{Chael}},
  \bibinfo{journal}{Phys. Rev. D} \textbf{\bibinfo{volume}{97}},
  \bibinfo{pages}{103530} (\bibinfo{year}{2018}), \eprint{1802.06788}.

\bibitem[{\citenamefont{Ooba et~al.}(2019)\citenamefont{Ooba, Tashiro, and
  Kadota}}]{Ooba:2019erm}
\bibinfo{author}{\bibfnamefont{J.}~\bibnamefont{Ooba}},
  \bibinfo{author}{\bibfnamefont{H.}~\bibnamefont{Tashiro}}, \bibnamefont{and}
  \bibinfo{author}{\bibfnamefont{K.}~\bibnamefont{Kadota}},
  \bibinfo{journal}{JCAP} \textbf{\bibinfo{volume}{09}}, \bibinfo{pages}{020}
  (\bibinfo{year}{2019}), \eprint{1902.00826}.

\bibitem[{\citenamefont{Essig et~al.}(2012{\natexlab{b}})\citenamefont{Essig,
  Manalaysay, Mardon, Sorensen, and Volansky}}]{Essig:2012yx}
\bibinfo{author}{\bibfnamefont{R.}~\bibnamefont{Essig}},
  \bibinfo{author}{\bibfnamefont{A.}~\bibnamefont{Manalaysay}},
  \bibinfo{author}{\bibfnamefont{J.}~\bibnamefont{Mardon}},
  \bibinfo{author}{\bibfnamefont{P.}~\bibnamefont{Sorensen}}, \bibnamefont{and}
  \bibinfo{author}{\bibfnamefont{T.}~\bibnamefont{Volansky}},
  \bibinfo{journal}{Phys. Rev. Lett.} \textbf{\bibinfo{volume}{109}},
  \bibinfo{pages}{021301} (\bibinfo{year}{2012}{\natexlab{b}}),
  \eprint{1206.2644}.

\bibitem[{\citenamefont{Roberts et~al.}(2016)\citenamefont{Roberts, Dzuba,
  Flambaum, Pospelov, and Stadnik}}]{Roberts:2016xfw}
\bibinfo{author}{\bibfnamefont{B.}~\bibnamefont{Roberts}},
  \bibinfo{author}{\bibfnamefont{V.}~\bibnamefont{Dzuba}},
  \bibinfo{author}{\bibfnamefont{V.}~\bibnamefont{Flambaum}},
  \bibinfo{author}{\bibfnamefont{M.}~\bibnamefont{Pospelov}}, \bibnamefont{and}
  \bibinfo{author}{\bibfnamefont{Y.}~\bibnamefont{Stadnik}},
  \bibinfo{journal}{Phys. Rev. D} \textbf{\bibinfo{volume}{93}},
  \bibinfo{pages}{115037} (\bibinfo{year}{2016}), \eprint{1604.04559}.

\bibitem[{\citenamefont{Catena et~al.}(2020)\citenamefont{Catena, Emken,
  Spaldin, and Tarantino}}]{Catena:2019gfa}
\bibinfo{author}{\bibfnamefont{R.}~\bibnamefont{Catena}},
  \bibinfo{author}{\bibfnamefont{T.}~\bibnamefont{Emken}},
  \bibinfo{author}{\bibfnamefont{N.~A.} \bibnamefont{Spaldin}},
  \bibnamefont{and}
  \bibinfo{author}{\bibfnamefont{W.}~\bibnamefont{Tarantino}},
  \bibinfo{journal}{Phys. Rev. Res.} \textbf{\bibinfo{volume}{2}},
  \bibinfo{pages}{033195} (\bibinfo{year}{2020}), \eprint{1912.08204}.

\bibitem[{\citenamefont{Aprile et~al.}(2019)}]{Aprile:2019xxb}
\bibinfo{author}{\bibfnamefont{E.}~\bibnamefont{Aprile}} \bibnamefont{et~al.}
  (\bibinfo{collaboration}{XENON}), \bibinfo{journal}{Phys. Rev. Lett.}
  \textbf{\bibinfo{volume}{123}}, \bibinfo{pages}{251801}
  (\bibinfo{year}{2019}), \eprint{1907.11485}.

\bibitem[{\citenamefont{Collaboration}(2020)}]{xenon:collaboration}
\bibinfo{author}{\bibfnamefont{X.}~\bibnamefont{Collaboration}},
  \emph{\bibinfo{title}{{XENON1T/s2only\_data\_release: XENON1T S2-only data
  release}}} (\bibinfo{year}{2020}),
  \urlprefix\url{https://doi.org/10.5281/zenodo.4075018}.

\bibitem[{\citenamefont{Berlin and Blinov}(2018)}]{Berlin:2017ftj}
\bibinfo{author}{\bibfnamefont{A.}~\bibnamefont{Berlin}} \bibnamefont{and}
  \bibinfo{author}{\bibfnamefont{N.}~\bibnamefont{Blinov}},
  \bibinfo{journal}{Phys. Rev. Lett.} \textbf{\bibinfo{volume}{120}},
  \bibinfo{pages}{021801} (\bibinfo{year}{2018}), \eprint{1706.07046}.

\end{thebibliography}

\onecolumngrid
\clearpage

\setcounter{page}{1}
\setcounter{equation}{0}
\setcounter{figure}{0}
\setcounter{table}{0}
\setcounter{section}{0}
\setcounter{subsection}{0}
\renewcommand{\theequation}{S.\arabic{equation}}
\renewcommand{\thefigure}{S\arabic{figure}}
\renewcommand{\thetable}{S\arabic{table}}
\renewcommand{\thesection}{\Roman{section}}
\renewcommand{\thesubsection}{\Alph{subsection}}
\newcommand{\ssection}[1]{
    \addtocounter{section}{1}
    \section{\thesection.~~~#1}
    \addtocounter{section}{-1}
    \refstepcounter{section}
}
\newcommand{\ssubsection}[1]{
    \addtocounter{subsection}{1}
    \subsection{\thesubsection.~~~#1}
    \addtocounter{subsection}{-1}
    \refstepcounter{subsection}
}
\newcommand{\fakeaffil}[2]{$^{#1}$\textit{#2}\\}

\thispagestyle{empty}
\begin{center}
    \begin{spacing}{1.2}
        \textbf{\large 
            \hypertarget{sm}{Supplemental Material:} \\ New Strong Bounds on sub-GeV Dark Matter from Boosted and Migdal Effect
        }
    \end{spacing}
    \par\smallskip
    Victor V. Flambaum,$^{1,2}$
    Liangliang Su,$^{3,4}$
    Lei Wu,$^{3}$
    and Bin Zhu,$^{4}$
    \par
    {\small
        \fakeaffil{1}{School of Physics, University of New South Wales, Sydney 2052, Australia}
        \fakeaffil{2}{Johannes Gutenberg-Universität Mainz, 55128 Mainz, Germany and Helmholtz-Institut, GSI Helmholtzzentrum für Schwerionenforschung, 55128 Mainz, Germany}
        \fakeaffil{3}{Department of Physics and Institute of Theoretical Physics, Nanjing Normal University, Nanjing, 210023, China} 
        \fakeaffil{4}{Department of Physics, Yantai University, Yantai 264005, China}
    }

\end{center}
\par\smallskip

In this supplemental material, we show the numerical results of the dark matter form factor $F_{\mathrm{DM}}$, the ionization form factor $f^{nl}_{ion}$, and the nucleus form factor $F_{N}$ in Sec.~\ref{ssec1}. We discuss the inverse speed function $\eta$ in Sec.~\ref{ssec2} and the attenuation effect in Sec.~\ref{ssec3}. Finally, we present the procedure of deriving the exclusion limit with Xenon1T data in Sec.~\ref{ssec4}.

\ssection{Numerical Results of Form Factors}
\label{ssec1}

\begin{figure}[ht]
  \centering
\includegraphics[height=10cm,width=12cm]{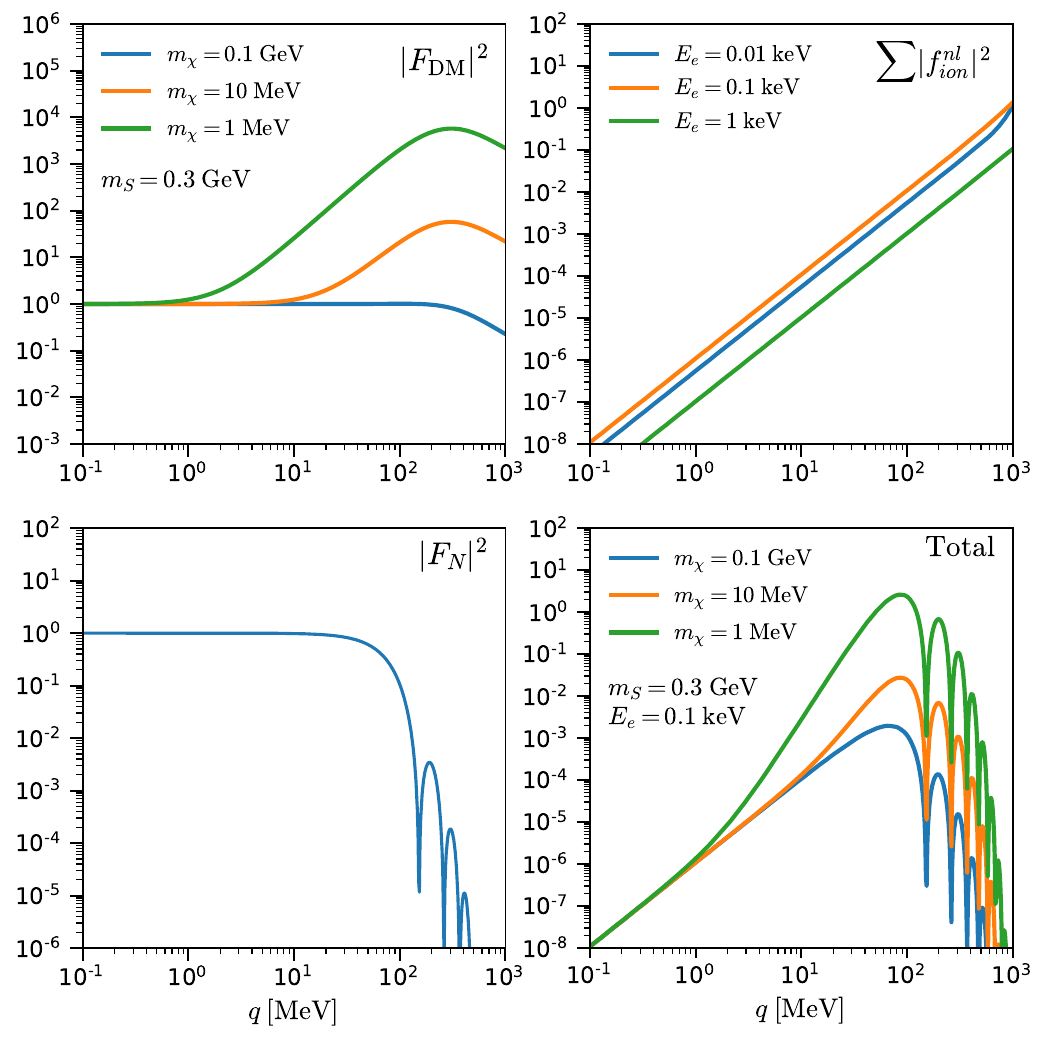}
\caption{The dependence of the DM form factor $\left|F_{\rm DM}\right|^2$ (top-left), the sum of the ionization factor $\sum\left|f^{nl}_{ion}\right|^2$ (top-right), the nucleus form factor $\left|F_N\right|^2$ (bottom-left), and their product (bottom-right) on the nucleus momentum transfer $q$. Note that the sum of ionization factor includes the contributions of $n=4,5$ shells in Xenon.}
\label{fig:formfactor}
\end{figure}

In this work, we study a simplified hadrophilic DM model, where a flavor-specific singlet scalar mediator $S$ couples to up-type quarks and DM. The relevant interactions are given in Eq.~\ref{lag}. With the effective coupling $g_{\mathbf n}$ between the scalar mediator and a nucleon, we can derive the squared matrix element of the DM-nucleon scattering as follows,
\begin{equation}
\left|\mathcal{M}\right|^{2}=\frac{g_{\mathbf n}^{2} g_{\chi}^{2}\left(t-4 m_{\mathbf n}^{2}\right)\left(t-4 m_{\chi}^{2}\right)}{\left(m_{S}^{2}-t\right)^{2}},
\end{equation}
where $t=-2m_N E_R=-q^2$ with $m_N$ being nucleus mass and $E_R$ being nucleus recoil energy. We define the squared matrix element at the reference momentum $q_0=\alpha m_e$ as 
\begin{equation}
\left|\mathcal{M}(q=q_0)\right|^2=\frac{g_{n}^{2} g_{\chi}^{2}\left(4 m_{n}^{2} + q^2_0 \right)\left( 4 m_{\chi}^{2}+q^2_0\right)}{\left(m_{S}^{2}+q^2_0\right)^{2}}.
\end{equation}
In terms of the definition of dark matter form factor in Eq.~\ref{formDM}, we obtain
\begin{equation}
\left|F_{\mathrm{DM}}(q)\right|^2=\frac{\left(4m_n^2+q^2\right) \left(m_S^2+q_0^2\right)^2 \left(4 m_{\chi}^2+q^2\right)}{ (4m_n^2+q_0^2) (4m_{\chi }^2+q_0^2) \left(m_S^2+q^2\right)^2}.
\label{formADM}
\end{equation}
In regions of our interested DM mass, $m_n \gg q_0$ and $m_{\chi} \gg q_0$, we can simplify the $|F_{\mathrm{DM}}|^2$ to,
\begin{equation}
\left|F_{\mathrm{DM}}(q)\right|^2=\frac{\left(4m_n^2+q^2\right) \left(m_S^2+q_0^2\right)^2 \left(4 m_{\chi}^2+q^2\right)}{ 16 m_n^2 m_{\chi }^2\left(m_S^2+q^2\right)^2}.
\end{equation}
In Fig.~\ref{fig:formfactor}, we show the dependence of the DM form factor $|F_{\rm DM}|^2$, the ionization factor $|f^{nl}_{ion}|^2$, the nucleus form factor $|F_N|^2$, and their product on the nucleus momentum transfer $q$. From the top-left panel of Fig.~\ref{fig:formfactor}, we can see that $\left|F_{\rm DM}\right|^2$ depends on the masses of the DM. For instance, the value of $\left|F_{\rm DM}\right|^2$ for $m_\chi=1$ MeV is about ${\cal O}(10^3)$ larger than that for $m_\chi=0.1$ GeV at $q \simeq 100$ MeV. In the top-right panel of Fig.~\ref{fig:formfactor}, it can be seen that the $|f_{ion}^{nl}|^2$ is enhanced by about eight orders when $q$ is varied from about 0.1 MeV to 1 GeV. Besides, from the top-right panel of Fig.~\ref{fig:formfactor}, we can find that the product of all factors decrease rapidly in the range of $q \gtrsim 100$ MeV. This is because that the nucleus form factor $\left|F_{N}\right|^2$ is highly suppressed in the range of $q \gtrsim 100$ MeV.

\ssection{Inverse Speed Function}
\label{ssec2}

We first consider the DM differential flux $\mathrm{d}\Phi_{\chi}=\mathrm{d}N_{\chi}/At$ in $z$-axis. Here $\mathrm{d}N_{\chi}$ is the number of DM with the velocity in the interval $[v_z, v_z+\mathrm{d}v_z]$ through the area $A$ over the time $t$,
\begin{equation}
   \mathrm{d} N_{\chi}=A v_z t \cdot n_{\chi}\cdot f(v_z)\mathrm{d}v_z,
\end{equation}
where $n_\chi=N_\chi/V$ is the number density of particles in the gas and $f(v_z)\mathrm{d}v_z$ is the number of the DM in the interval $[v_z,v_z+\mathrm{d}v_z]$. Then we can have the DM differential flux as following,
\begin{equation}
    \mathrm{d}\Phi_{\chi}=n_{\chi}v_z f(v_z)\mathrm{d}v_z.
\end{equation}
It is straightforward to generalize the above differential flux into $3$-dimensional case,
\begin{equation}
\mathrm{d}\Phi_{\chi}=n_{\chi}|\mathbf{v}|f(\mathbf{v})\mathrm{d}^3\mathbf{v}=n_{\chi} v^3 f(v)\sin\theta \mathrm{d}\theta \mathrm{d}\phi \mathrm{d} v.
    \end{equation}
With this result, we can calculate the inverse speed function $\eta$, 
\begin{eqnarray}
\eta\left(E_{\min }\right)&=&\int_{E_{\min }} \mathrm{d} E_{\chi} n_{\text {halo }}^{-1} \frac{m_{\chi}^{2}}{p_{\chi}^2} \frac{\mathrm{d} \Phi_{\chi}}{\mathrm{d} E_{\chi}}=\int_{E_{\min }}\left(\frac{1}{n_{\text {halo }}}\right) \frac{m_{\chi}^{2}}{v^2 E_{\chi}^{2}} n_{\chi} v f(\mathbf v) \mathrm{d}^{3} \mathbf v \nonumber \\
&=&\left(\frac{n_{\chi}}{n_{\text {halo }}}\right)\int_{E_{\min }} \frac{m_{\chi}^{2}}{v E_{\chi}^{2}} f(\mathbf v) \mathrm{d}^{3} \mathbf v,  
\end{eqnarray}
where $p_{\chi}/E_{\chi}=v$ is used. For the halo DM, we have $E_{\chi}=m_{\chi}$ and $n_{\chi} =n_{\text {halo }} $ so that the $\eta$ function can be reduced to the standard halo dark matter $\eta$ function, 
\begin{equation}    \eta(v_{\min})=\int_{v_{\min}}\frac{f(\mathbf{v})}{v}\mathrm{d}^3 \mathbf v,
\end{equation}
where $f(\mathbf{v})$ is the velocity function that obeys Maxwell-Boltzmann distribution. For the boosted dark matter, like our ADM, the $\eta$ function becomes
\begin{equation}
\eta\left(E_{\chi}^{\min }\right)=\int_{E_{\chi}^{\min }} \mathrm{d} E_{\chi} n_{\text {halo }}^{-1} \frac{m_{\chi}^{2}}{p_{\chi}^2} \frac{\mathrm{d} \Phi_{\chi}}{\mathrm{d} E_{\chi}},
\end{equation}
where $\mathrm{d} \Phi_{\chi}/\mathrm{d} E_{\chi}=\mathrm{d} \Phi_{\chi}/\mathrm{d} T_{\chi}$ is the flux of the ADM at the detector that will be discussed in the next section.

\ssection{Attenuation Effect}
\label{ssec3}

The differential flux of the ADM at the Earth's surface is given by,
\begin{eqnarray}
\frac{\mathrm{d} \phi_{\chi}}{\mathrm{d} T_{\chi}^0}=G_0 \int_{T_{p}^{\mathrm{min}}}^{T_{p}^{\max }}  \frac{\mathrm{d} T_{p}}{\Omega\left(T_{p}\right)} \frac{\mathrm{d}\phi_{p}\left(h_{\mathrm{max}}\right)}{\mathrm{d} T_{p}} \frac{\mathrm{d} \sigma_{p N \to \chi \chi X}}{\mathrm{d} T_{\chi}},
\end{eqnarray}
where the differential flux of the CRs inside the atmosphere $ \mathrm{d} \phi_{p}/\mathrm{d} T_{p}$ is a function of the height $h$ from the ground level. The inelastic differential cross section of the proton-nitrogen scattering can be approximately calculated by $d\sigma_{pN \to \pi \chi \bar{\chi}}/dT_{\chi}\simeq \frac{\sigma_{pN}}{T_{\chi}^{\max}} \mathrm{BR}(\eta \to  \chi\chi X)$. In our numerical calculations, we use the package {\textsf{CRMC}}~\footnote{See the CRMC package:https://doi.org/10.5281/zenodo.4558706 and Alvey's code in the github:https://github.com/james-alvey-
42/BoostedDM. } to simulate this process and obtain $\sigma_{pN} \simeq 255$ mb. The geometrical factor $G_0$ is given by,
\begin{eqnarray}
G_0&=&2\pi\int_{0}^{h_{\mathrm{max}}} \mathrm{d} h \int_{-1}^{+1} \mathrm{d}\cos\theta \left(R_{E}+h\right)^{2} \frac{y_{p}}{\ell_{d}^{2}} n_{N}(h) ,
\end{eqnarray}
where $n_N$ is the number density of the nitrogen. The line of the sight distance between the production point of DM and the detector is $\ell_{d}^{2}=\left(R_{E}+h\right)^{2}+\left(R_{E}-h_{d}\right)^{2} -2\left(R_{E}+h\right)\left(R_{E}-h_{d}\right) \cos \theta$, where $h_d=1.4$ km is the depth of the detector and $\theta$ is the polar angle of the DM production point. The incoming CRs are diluted in the atmosphere and the corresponding dilution factor $y_p$ is given by $y_p=\exp \left(-\sigma_{p N} \int_{h}^{h \max } \mathrm{d} h' n_{N}(h')\right)$, where $h_{\max}=180$ km and $R_E=6378.1$ km. Assuming the Earth is transparent, we checked and found the geometrical factor $G$ in Ref.~\cite{Alvey:2019zaa} will return to the factor $G_0$. Before reaching the detector, the DM will travel through the Earth and then be attenuated by the interactions with the ordinary matter. During this process, the energy loss of the ADM can be calculated by 
\begin{equation}
\frac{\mathrm{d} T_{\mathrm{\chi}}}{\mathrm{d} l}=-\sum_{i} n_{i}(\vec{r}) \int_{0}^{E_{R}^{\max }} \frac{\mathrm{d} \sigma_{\chi i}}{\mathrm{d} E_{R}} E_{R} \mathrm{d} E_{R}.
\label{loss}
\end{equation}
where $n_i(\vec{r})$ is the number density of the different medium particles $i$ of the Earth at position $\vec{r}$. In the rest frame of Earth, the maximum recoil energy of the DM-nucleus scatting $E_{R}^{\max}$ for the boosted DM can be given by
\begin{equation}
E_{R}^{\max }=\frac{T_{\chi}^{2}+2 m_{\chi} T_{\chi}}{T_{\chi}+\left(m_{i}+m_{\chi}\right)^{2} /\left(2 m_{i}\right)}, 
\end{equation}
where $T_{\chi}$ is the kinetic energy of the DM. The differential cross section $\mathrm{d} \sigma_{\chi i}/\mathrm{d} E_{R}$ in our model is given by 
\begin{equation}
\begin{aligned}
\frac{\mathrm{d} \sigma_{\chi i}}{\mathrm{d} E_{R}}&=\frac{\bar{\sigma}_{n} m_{i} m_{\chi}^2 A_i^2}{2 \mu_{n}^{2} T_{\chi}(2m_{\chi}+T_{\chi})}\left|F_{\mathrm{DM}}(E_R)\right|^{2}\left|F_{N}(E_R)\right|^{2},
\end{aligned}
\end{equation} 
Then, the energy loss function in Eq.~\ref{loss} can be rewritten as   
\begin{eqnarray}
\frac{\mathrm{d} T_{\mathrm{\chi}}}{\mathrm{d} l}&=&-\sum_{i} n_{N}(\vec{r}) \frac{\bar{\sigma}_{n} m_{i}}{2 \mu_{n}^{2}}\frac{m_{\chi}^2}{T_{\chi}^2+2 m_{\chi} T_{\chi}}A_i^{2}  \times \int_{0}^{E_{R}^{\max }}   \left|F_{\mathrm{DM}}(E_R)\right|^{2}\left|F_{N}(E_R)\right|^{2} E_{R} \mathrm{d} E_{R} \nonumber \\
&=&-\sum_{i} n_{i}(\vec{r}) \frac{\bar{\sigma}_{n} m_{i}}{2 \mu_{n}^{2}}\frac{m_{\chi}^2}{T_{\chi}^2+2 m_{\chi} T_{\chi}}A_i^{2}  \times (E_{R}^{\max })^2\int_{0}^{1} x \left|F_{\mathrm{DM}}(x E_R^{\max})\right|^{2}\left|F_{N}(x E_R^{\max})\right|^{2} \mathrm{d} x \label{eq:Tchi}
\end{eqnarray}
By solving this differential equation numerically, we can obtain the kinetic energy of the ADM $T_{\chi}$ at the detector. Finally, we can also calculate the ratio $\mathrm{d} T_{\chi}^0/\mathrm{d} T_{\chi}$ and the flux of the ADM at the detector,
\begin{equation}
    \frac{\mathrm{d} \Phi_{\chi}}{\mathrm{d} T_{\chi}} = \int \mathrm{d} \Omega \frac{\mathrm{d} T_{\chi}^0}{\mathrm{d} T_{\chi}} \frac{\mathrm{d} \Phi}{\mathrm{d} \Omega \mathrm{d} T_{\chi}^0}  = 2 \pi \int_{0}^{\pi} \mathrm{d} \theta \sin{\theta} \frac{\mathrm{d} T_{\chi}^0}{\mathrm{d} T_{\chi}} \frac{\mathrm{d} \Phi}{\mathrm{d} \Omega \mathrm{d} T_{\chi}^0}
\end{equation}

\ssection{Using Xenon1T data to set the limits}
\label{ssec4}

{\it Nucleus Recoil Data.} We calculate the event rate of the ADM-nucleus scattering in the detector as follows,
\begin{equation}
R_{\mathrm{ADM}}=\int_{E_1}^{E_2}\mathrm{d}E_R\frac{\mathrm{d} R}{\mathrm{d} E_{R}}=\int_{E_1}^{E_2}\mathrm{d}E_R N_T \int_{T_{\chi}^{\min }}^{\infty} \mathrm{d} T_{\chi} \frac{\mathrm{d} \Phi_{\chi}}{\mathrm{d} T_{\chi}^z} \frac{\mathrm{d} \sigma_{\chi N}}{\mathrm{d} E_{R}},
\label{rate1}
\end{equation}
where we focus on the nucleus recoil energy $E_R$ in the range of $[4.9, 40.9]$ keV. On the other hand, the experimental data can be recast as a total event rate,
\begin{equation}
\begin{aligned}
    R_{\mathrm{Xenon}}=N_T\kappa \frac{\sigma_{\chi N}^{\mathrm{DM}}}{m_{\mathrm{DM}}}
    \left(\bar{v} \rho_{\mathrm{DM}}\right)^{\text {local }},
    \end{aligned}
    \label{rate2}
\end{equation}
where $\kappa=0.23$ and 
$\sigma_{\chi N}^{\mathrm{DM}}/m_{\mathrm{DM}} \simeq 8.3\times 10^{-49}\mathrm{cm}^2/\mathrm{GeV}$~\cite{Bringmann:2018cvk}. Comparing Eq.~\ref{rate1} with~\ref{rate2}, we can derive the exclusion limits of the ADM-nucleus scattering cross section.


{\it Electron Recoil Data.} To compare with S2 data, we calculate the different event rate of the ionized electron $\mathrm{d} R_{i o n} / \mathrm{d} E_{e}$ as,
\begin{equation}
\frac{\mathrm{d} R_{ion}}{\mathrm{d}{\rm S2}}= \int d E_{e} \epsilon({\rm S2}) P\left({\rm S2} \mid E_{\mathrm{EM}}\right) \mathrm{d} R_{ion} / \mathrm{d} E_{e},
\end{equation}
where $\epsilon({\rm S2})$ and $P\left({\rm S2} \mid E_{\mathrm{EM}}\right)$ are the detector efficiency and the probability function that converts deposited electromagnetic energy $E_{\mathrm{EM}}=|E_{nl}|+ E_e$ into the photoelectron (PE), respectively. Their product is the so called the response function, which is tabulated by XENON1T collaboration in Ref.~\cite{Aprile:2019xxb, xenon:collaboration}. In our work, we are interested in the region with the electron recoil energy greater than $0.186$ keV for the Migdal effect because the lower energy region is assumed undetectable in Ref.~\cite{Aprile:2019xxb}. We utilize the differential ionization rate multiplying the response function matrix to calculate the $\mathrm{d} R_{ion}/{\mathrm{d}{\rm S2}}$ in different regions of interest (ROIs). Finally, we obtain the limits on our model by comparing the theoretical predictions with the experimental data provided in Ref.~\cite{xenon:collaboration}. 

\end{document}